\documentclass[aps,prb,superscriptaddress,twocolumn,floatfix,showpacs,citeautoscript,longbibliography,hyperlinks]{revtex4-1}
\usepackage{amsmath}
\usepackage{amssymb}
\usepackage{amsfonts}
\usepackage{dsfont}
\usepackage{color}
\usepackage{nicefrac}
\usepackage[autostyle]{csquotes}
\usepackage[colorlinks=true,citecolor=blue]{hyperref}
\usepackage{graphicx,graphics}
\usepackage{environ}
\usepackage{siunitx}
\usepackage[dvipsnames]{xcolor}
\usepackage{xcolor}
\usepackage{comment}

\usepackage{soul}

\begin{document}

\title{Real-time observation of Cooper pair splitting showing strong non-local correlations}

\author{Antti Ranni}
\email{antti.ranni@ftf.lth.se}
\affiliation{NanoLund and Solid State Physics, Lund University, Box 118, 22100 Lund, Sweden}
\author{Fredrik Brange}
\author{Elsa T. Mannila}
\author{Christian Flindt}
\affiliation{Department of Applied Physics, Aalto University, 00076 Aalto, Finland}
\author{Ville F. Maisi}
\email{ville.maisi@ftf.lth.se}
\affiliation{NanoLund and Solid State Physics, Lund University, Box 118, 22100 Lund, Sweden}

\date{\today}

\maketitle

{\bf Controlled generation and detection of quantum entanglement between spatially separated particles constitute an essential prerequisite both for testing the foundations of quantum mechanics and for realizing future quantum technologies. Splitting of Cooper pairs from a superconductor provides entangled electrons at separate locations. However, experimentally accessing the individual split Cooper pairs constitutes a major unresolved issue as they mix together with electrons from competing processes. Here, we overcome this challenge with the first real-time observation of the splitting of individual Cooper pairs, enabling direct access to the time-resolved statistics of Cooper pair splitting. We determine the correlation statistics arising from two-electron processes and find a pronounced peak that is two orders of magnitude larger than the background. Our experiment thereby allows to unambiguously pinpoint and select split Cooper pairs with 99 \% fidelity. These results open up an avenue for performing experiments that tap into the spin-entanglement of split Cooper pairs.}

Cooper pair splitters are promising solid-state devices for generating non-local entanglement of electronic spins.\cite{Recher2001,Lesovik2001} The basic operating principle is based on the tunneling of spin-entangled electrons from a superconductor into spatially separated normal-state structures, whereby entanglement between different physical locations is obtained. The controlled generation of entangled particles is not only of fundamental interest to test the foundations of quantum mechanics.\cite{Brunner2014} It is also a critical prerequisite for future quantum technologies,\cite{zagoskin2011,shevchenko2019} such as quantum information processors and other quantum devices, which will be `fuelled' by entanglement.

Until now, the splitting of Cooper pairs has been indirectly inferred from measurements of the currents in the outputs of a Cooper pair splitter\cite{Beckmann2004,Russo2005,Yeyati2007,Hofstetter2009,Zimansky2009,Herrmann2010,Wei2010,Hofstetter2011,Herrmann2012,Das2012,Fueloep2014,Schindele2014,Tan2015,Fulop2015,Borzenets2016,Schindele2012,deacon2015,Ueda2019,tan2020,Bruhat2018,Pandey2021} or their low-frequency cross-correlations.\cite{Wei2010,Das2012} These approaches rely on measuring the average currents, or small fluctuations around them, due to a large number of splitting events. The currents consist of contributions from Cooper pair splitting as well as other competing transport processes, which not only makes these experiments highly challenging, but also hinders direct access to the correlated electron pairs. In particular, a considerable fraction of the electrical signal is due to unwanted processes, and those electrons that actually are  entangled have already passed through the device and are lost, once the currents have been measured. 

Here we take a fundamentally different route and use charge detectors to observe the splitting of individual Cooper pairs in real time as it happens. Unlike earlier experiments, we use isolated islands that are not connected to external drain electrodes. Hence, once a Cooper pair is split, the two correlated electrons remain stored on the islands and can be detected in real-time.

\section*{Results}

\begin{figure*}[t]
	\centering
	\includegraphics[width=180mm]{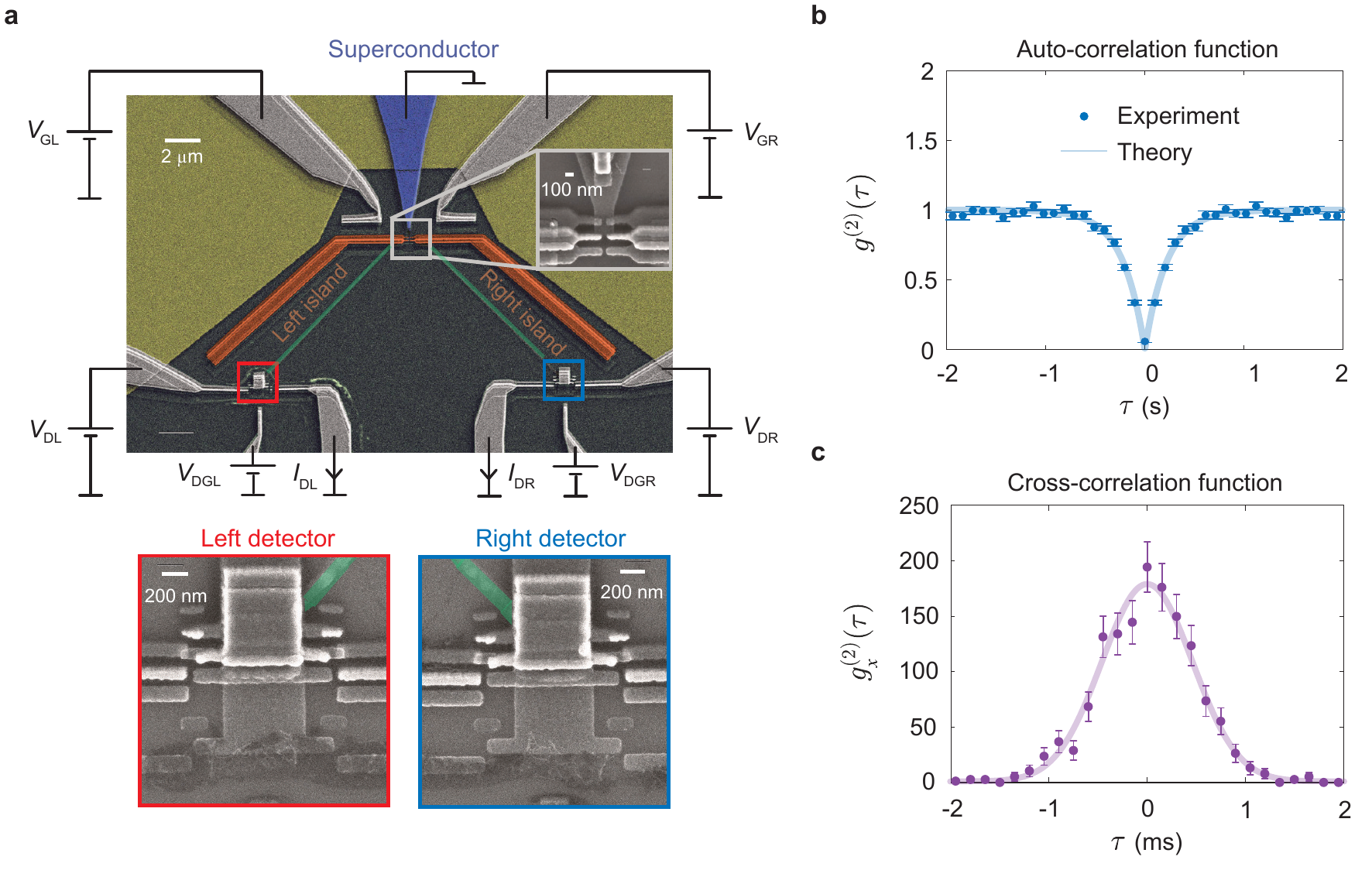}
	\caption{\textbf{Cooper pair splitter and correlation measurements.}
	\textbf{a,} Scanning electron micrograph of the grounded superconducting reservoir (colored blue) coupled via tunnel junctions to two normal-state metallic islands (in orange). The voltages $V_{\mathrm{GL}}$ and $V_{\mathrm{GR}}$ are connected to gate electrodes and tuned to control the electronic populations on the islands. Each island is capacitively coupled via a metal strip (in green) to a single-electron transistor, which serves as a real-time charge detector to read out the number of electrons on each island from the measured signals $I_{\mathrm{DL}}$ and $I_{\mathrm{DR}}$. The detectors are biased by the voltages $V_{\mathrm{D}\alpha}$ ($\alpha = L$ or $R$) and gated by the voltages $V_{\mathrm{DG}\alpha}$. The inset in the upper right corner shows the tunnel junctions that connect the superconductor to the two metallic islands.  
	\textbf{b,} Measurements of the auto-correlation function $g^{(2)}$ for single-electron tunneling into the right island with a time delay $\tau$ between tunneling events. The correlation function is suppressed at short times due to the strong Coulomb interactions on the island, which lead to anti-bunching.
	\textbf{c,} Cross-correlation function $g^{(2)}_x$ for transitions between the superconductor and the left island at the time $t=0$, followed by transitions between the superconductor and the right island at the time $t=\tau$. The cross-correlations are strongly enhanced for short times due to crossed two-electron processes such as Cooper pair splitting and elastic cotunneling between the islands via the superconductor. The error bars in \textbf{b} and \textbf{c} are given by the standard deviations. Details of the device fabrication and the measurements are provided in the Methods section. \label{fig1}
		}
\end{figure*}

{\bf Device architecture.} Figure~\ref{fig1}\textbf{a} shows a scanning electron microscope image of one of our Cooper pair splitters made of a superconducting aluminium electrode coupled to two normal-state copper islands. The grounded superconducting electrode and the islands are connected via insulating aluminium oxide layers acting as tunnel junctions. The superconductor functions as a source of Cooper pairs, which are split into two separate electrons that tunnel into the islands, one in each. We detect the number of electrons on each island using single-electron transistors whose conductance depends sensitively on the charge state of the islands.\cite{schoelkopf1998,Gustavsson2009,pekola2013} Each detector is biased by a voltage $V_{\mathrm{D\alpha}}$ ($\alpha = L$ or $R$) and the currents $I_{\mathrm{D\alpha}}$ through them are measured in the grounded contacts. The detectors are tuned to charge sensitive operation points by the gate voltages $V_{\mathrm{DG\alpha}}$. We can thus monitor the tunneling of electrons in and out of the islands, including simultaneous processes\cite{maisi2011,maisi2014}, such as Cooper pair splitting, where one electron tunnels into each of the two islands at the same time. The electronic population of each island is feedback-controlled by the gate voltages $V_{\mathrm{G\alpha}}$, which are adjusted to maintain the two lowest-lying charge states at equal energies.\cite{Koski2014,Singh2019} Hence, there is no energy cost for tunneling processes that change the electron number on each island by one.

{\bf Correlation measurements.} Figures~\ref{fig1}\textbf{b} and \textbf{c} show correlation measurements using our charge detectors. The $g^{(2)}$-function describes correlations between tunneling processes with a time delay $\tau$ between them and are for example used extensively in quantum optics to characterize light sources.\cite{Loudon2000} A value of $g^{(2)}(\tau) = 1$  at all times implies that the particles are uncorrelated. On the other hand, a $g^{(2)}$-function that peaks at short times indicates that the particles tend to bunch as for instance for thermal light, which has $g^{(2)}(0)=2$ at $\tau=0$. By contrast, a coherent single-photon source is characterized by a $g^{(2)}$-function with a dip at short times, indicating that the particles are anti-bunched, ideally with $g^{(2)}(0)=0$.

Here we measure the auto-correlations for tunneling events from the superconductor to the right island as well as the cross-correlations for tunneling between the superconductor and each of the two islands. The auto-correlations in Fig.~\ref{fig1}\textbf{b} are fully suppressed at short times. The suppression arises from the strong Coulomb interactions, which prevent more than one electron at a time to tunnel into the right island, leading to anti-bunching of the tunneling events. A theoretical analysis captures the experimental findings in Fig.~\ref{fig1}\textbf{b} by the expression
\begin{equation}
    g^{(2)}(\tau)=1-e^{-\gamma |\tau|},
    \label{Auto g2}
\end{equation}
where $\gamma= 4.5$~s$^{-1}$ is the inverse correlation time, which may be determined from the tunneling rates (see Supplementary Note 2 for details). Equation (\ref{Auto g2}) is similar to the auto-correlation function of a simple two-level system, however, in our case, it follows from an elaborate model of the Cooper pair splitter, which involves several charge states of each island, and the inverse correlation time is a complicated function of all tunneling rates.

\begin{figure*}[t]
	\centering
	\includegraphics[width=180mm]{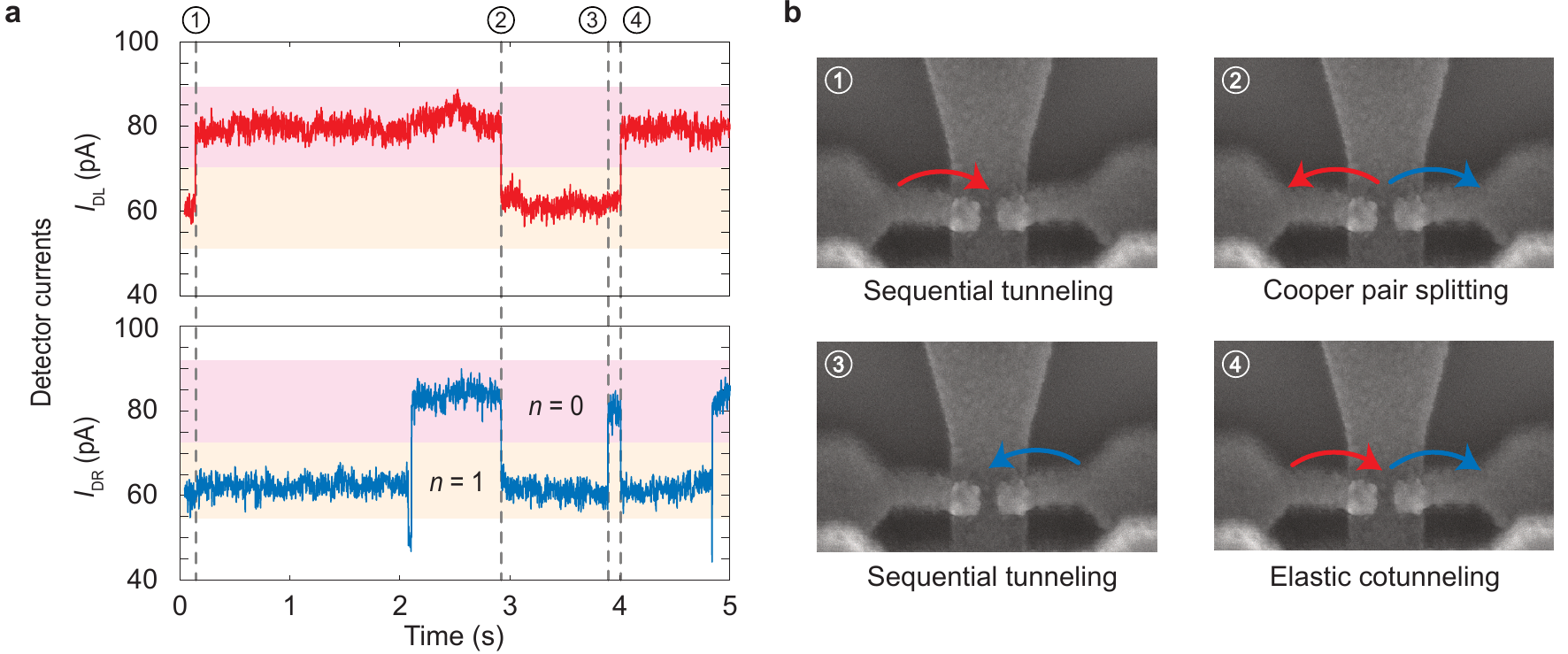}
	\caption{\textbf{Real-time observation of Cooper pair splitting.} \textbf{a,} Typical time traces of the currents $I_{\mathrm{DL}}$ and $I_{\mathrm{DR}}$ in the left and the right single-electron transistors correspondingly, which switch between two distinct levels corresponding to having $n=0$ or 1 (excess) electrons on each island. \textbf{b,} The points marked with \raisebox{.5pt}{\textcircled{\raisebox{-.9pt} {1}}} and \raisebox{.5pt}{\textcircled{\raisebox{-.9pt} {3}}} in the left panel correspond to processes, where a single electron tunnels between the superconductor and one of the islands. The point marked with \raisebox{.5pt}{\textcircled{\raisebox{-.9pt} {2}}} is a Cooper pair splitting event, where two electrons simultaneously tunnel from the superconductor into the two islands, one in each,  while the point marked with \raisebox{.5pt}{\textcircled{\raisebox{-.9pt} {4}}} is an elastic cotunneling event, where an electron is transferred between the two islands via the superconductor.}
		\label{fig2}
\end{figure*}

Turning next to the cross-correlations in Fig.~\ref{fig1}\textbf{c}, a completely different picture emerges. We now consider the conditional probability of observing a tunneling event between the superconductor and the right island at the time delay $\tau$ after a tunneling event between the superconductor and the left island has occurred. In this case, we observe a large peak at short times which is two orders of magnitude larger than the uncorrelated background of $g^{(2)}_x(\tau) = 1$. These correlations are a direct manifestation of nearly instantaneous two-electron processes involving both islands. The two-electron processes lead to correlated single-electron events in the two islands occurring on a microsecond timescale, which is much faster than the correlation time of the single-electron processes for each island as seen in Fig.~\ref{fig1}\textbf{b}. Theoretically, we can describe the cross-correlations as
\begin{equation}
    g^{(2)}_x(\tau) = 1+ \alpha_2 \frac{e^{-\frac{1}{2}(\tau/\sigma_D)^2}}{\sqrt{2\pi}\sigma_D},
    \label{Cross g2}
\end{equation}
where $\sigma_D = 460$~$\mu$s is the broadening due to timing jitter of the detectors, and $\alpha_2 = 210$~ms is the time-integrated contribution from two-electron processes. Figure~\ref{fig1}\textbf{c} shows that this expression agrees well with the experiment, explaining the strong non-local correlations. Figure~\ref{fig1} also illustrates how the cross-correlations, unlike the auto-correlations, can be measured on a time-scale, which is shorter than the rise time of each detector of about 4 ms: The cross-correlations concern tunneling events in different islands and are not limited by the dead time of the detectors (see Supplementary Note 1).

{\bf Cooper pair splitting in real-time.} We now address the identification of the individual tunneling events. Typical time-traces of the electrical currents in the two single-electron transistors are shown in Fig.~\ref{fig2}\textbf{a}, illustrating how we can detect the tunneling events in each island. At the time marked by \raisebox{.5pt}{\textcircled{\raisebox{-.9pt} {1}}}, the current in the left detector (red curve) switches suddenly from 60 pA to 80 pA as an electron tunnels out of the left island as depicted in panel \textbf{b}. At the time marked by \raisebox{.5pt}{\textcircled{\raisebox{-.9pt} {2}}}, the current in the left detector switches again as an electron tunnels into the left island. However, this time, the current in the right detector (blue curve) also changes as an electron tunnels into the right island. As we discuss below, statistical arguments allow us to conclude with near unity probability that these simultaneous tunneling events occur due to the splitting of a Cooper pair as indicated in panel \textbf{b}, rather than being two uncorrelated single-electron processes. Continuing further in time, the point marked by~\raisebox{.5pt}{\textcircled{\raisebox{-.9pt} {3}}} again corresponds to the tunneling of a single electron as in \raisebox{.5pt}{\textcircled{\raisebox{-.9pt} {1}}}, except that now an electron tunnels out of the right island. Finally, at the time marked by \raisebox{.5pt}{\textcircled{\raisebox{-.9pt} {4}}}, the detector currents switch in opposite directions due to elastic cotunneling in which an electron tunnels from the left to the right island via the superconductor.  

\begin{figure*}[t]
	\centering
	\includegraphics[width=180mm]{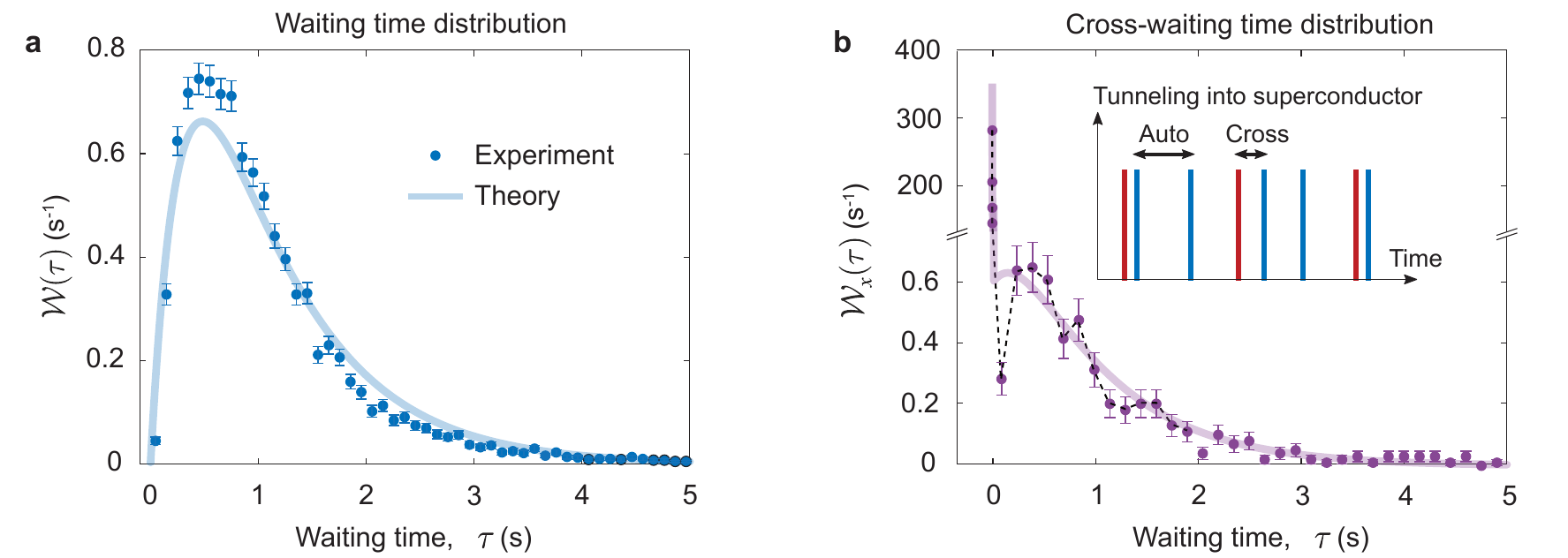}
	\caption{{\bf Waiting time distributions $\mathcal{W}$.} 
	\textbf{a,} Waiting times $\tau$ between single-electron tunneling out of the right island. The theory curve given by Eq.~(\ref{Auto-WTD right island}) agrees well with the measurements. \textbf{b,} Waiting times $\tau$ between single-electron tunneling out of the left island followed by single-electron tunneling out of the right island. Note that the peak at the short times is more than two orders of magnitude higher than the long time data. The experimental data is well captured by Eq.~(\ref{cross-WTD}). The dashed line serves as a guide to the eye. The inset illustrates the waiting times between tunneling events of the same or different types.  The blue lines correspond to tunneling out of the right island, while the red ones correspond to the left island. The error bars mark $1\sigma$ confidence intervals.
		\label{fig3}}
\end{figure*}

We now return to the alleged Cooper pair splitting at the time marked by \raisebox{.5pt}{\textcircled{\raisebox{-.9pt} {2}}}. Strictly speaking, the two simultaneous tunneling events could in principle be completely uncorrelated. However, to rule out that scenario, we  again consider our cross-correlation measurements in Fig.~\ref{fig1}\textbf{c}. Uncorrelated single-electron processes give rise to a flat background with $g_x^{(2)}(\tau)=1$, while correlated two-electron processes, such as Cooper pair splitting, lead to the pronounced peak at short delay times, and they essentially all take place within a time window of width $ 6\sigma_D\simeq 3\ \mathrm{ms}$. Within this time window, the fraction of uncorrelated single-electron processes is $6\sigma_D/(\alpha_2+6\sigma_D)$, while the fraction of correlated two-electron processes is $\alpha_2/(\alpha_2+6\sigma_D)$. Hence, with fidelity $\mathcal{F}= \alpha_2/(\alpha_2+6\sigma_D) \simeq 99\%$ we can say that the point marked by \raisebox{.5pt}{\textcircled{\raisebox{-.9pt} {2}}} represents the splitting of a Cooper pair.

{\bf Waiting time distributions.} Based on the identification of the involved tunneling processes, we can determine all relevant tunneling rates in our experiment. Furthermore, additional information about the tunneling processes can be obtained from measurements of the electron waiting times.\cite{Walldorf2018,Walldorf2020,Wrzesniewski2020} Figure~\ref{fig3}\textbf{a} shows the distribution of waiting times between \textit{consecutive} electrons tunneling out of the right island. Unlike the correlation functions, this is an \textit{exclusive} probability density, since no tunneling events of the same type are allowed during the waiting time. Electron waiting times are typically hard to measure, since they require nearly perfect detectors, unlike the $g^{(2)}$-functions. The probability to observe a short waiting time is low, since only one electron at the time can tunnel out of the island. After having reached its maximum, the distribution falls off as it becomes exponentially unlikely to wait a long time for the island to be refilled and the next electron can tunnel out. These measurements agree well with the theoretical expectation
\begin{equation}
\mathcal{W}(\tau) = \frac{1}{\langle \tau \rangle u }\left(e^{-\gamma(1-u) \tau/2}-e^{-\gamma(1+u)\tau/2}\right),
\label{Auto-WTD right island}
\end{equation}
which in addition to the inverse correlation time from Eq.~\eqref{Auto g2} also contains the average waiting time between tunneling events, $\langle \tau \rangle = 1.2$~s, which enters through the parameter $u=\sqrt{1-4/(\gamma \langle\tau\rangle)}$. At short times, $\gamma\tau\ll1$, the waiting time distribution is linear in time, $\mathcal{W}(\tau)\simeq \gamma \tau/\langle \tau \rangle$, just as the $g^{(2)}$-function.  By contrast, at longer times, where the $g^{(2)}$-function flattens out, it describes the small probabilities to observe long waiting times.

Turning to the cross-waiting time distribution in Fig.~\ref{fig3}\textbf{b}, we consider here the waiting time between an electron tunneling out of the left island followed by the tunneling of an electron out of the right island. These experimental results are well-captured by a detailed theoretical analysis, which yields the expression
\begin{equation}
    \mathcal{W}_{x}(\tau) \!= \!\frac{\eta_0}{2}\left[ \sqrt{\frac{2}{\pi}}\frac{e^{-\frac{\tau^2}{2\sigma_D^2}}}{\sigma_D} \!+\!\mathcal{W}(\tau) \right]+(1-\eta_0)\mathcal{W}_0(\tau),
    \label{cross-WTD}
\end{equation}
where the first term in the brackets arises from the correlated two-electron processes, happening with the weight $\eta_0/2$, where $\eta_0\simeq 0.36$, such as processes where an electron from each island tunnels into the superconductor to form a Cooper pair. The bracket also contains the distribution $\mathcal{W}(\tau)$ from Equation~\eqref{Auto-WTD right island}, corresponding to the waiting time between a two-electron process and a subsequent transition on the right island. The last term in Equation~(\ref{cross-WTD}) originates from tunneling processes from the right island into the superconductor, which are weakly correlated with the left island and obey a Poissonian waiting time distribution $\mathcal{W}_0(\tau) \propto e^{-\gamma(1-u)\tau/2}$ for long times. Importantly, the exponential dependence at long times allows us to characterize the storage time of the split electron pairs on the islands, i.e., the typical duration from {\textcircled{\raisebox{-.9pt} {2}}} to {\textcircled{\raisebox{-.9pt} {3}}} in Fig.~\ref{fig2}\textbf{a}. We note that the storage time ($\sim\gamma^{-1}$) is more than two orders of magnitude longer than the detection time of the splitting process ($\sim \sigma_D$). 

\section*{Discussion}

We have observed the splitting of individual Cooper pairs in real-time and thereby enabled the identification and storage of single split Cooper pairs, a procedure which is not possible with conventional current and noise measurements. As such, our work paves the way for a wide range of future experimental developments. Specifically, by embedding the single-electron detectors in a radio-frequency circuit, it should be possible to read out the charge states in a few microseconds,\cite{schoelkopf1998,Bauerle_2018} i.e., on a timescale that is much shorter than the spin lifetime and coherence time of semiconductor quantum dots, which typically exceed hundred microseconds.\cite{hanson2007,zwanenburg2013} 
In addition to the fast detection, the other key ingredient to probe the entanglement of the electrons is to detect the spin states of the split electrons. The readout methods of spin qubits are based on either the energy splitting of the spin states or spin-blockade, which would readily provide the required techniques.\cite{hanson2007,zwanenburg2013} Alternatively, ferromagnetic leads may be used to determine the spin state of the electrons.\cite{Cottet2004,Cottet2004b} 
Thus, by combining the fast charge detection with the spin read-out, one may witness and certify the quantum entanglement of the split Cooper pairs.\cite{PhysRevB.89.125404,PhysRevLett.118.036804}  Ultimately, several Cooper pair splitters may be combined to realize the first rudimentary quantum information algorithms using split Cooper pairs. Interestingly, our real-time detection scheme of crossed events would also enable to discriminate different types of processes based on their statistics. For example, the instantaneous Cooper pair splitting events generate a sharp detection-limited peak as presented here, whereas the spin selective tunneling discussed by in Refs.~\onlinecite{Cottet2004,Cottet2004b} would yield tunneling time broadened peak that could be probed with the correlation function measurements. In conclusion, our work opens up avenues for experiments using entangled spins, and it thereby enables future quantum technologies based on entangled electrons in on-chip solid-state devices.

\section*{Methods}

{\bf Device fabrication.} The device was fabricated on a silicon substrate with a 300 nm thermal silicon oxide layer on top. The fabrication started with a 2/30/2 nm thick patterned Ti/Au/Ti layer (yellow in Fig.~\ref{fig1}\textbf{a}) acting as a ground plane to filter out stray high-frequency noise. Then, a 10 nm thick chromium strip (green line running in 45$^\circ$ angles in the figure) was patterned and deposited to obtain a capacitive coupling between the islands and the detectors. Next, a 40 nm thick AlOx layer was grown with atomic layer deposition to electrically insulate the ground plane and the coupler from the rest of the structures. Finally, electron beam lithography and shadow mask deposition were used to make the Cooper pair splitter and the charge detectors. 
In the last step, four metal layers were deposited at different angles with two oxidation steps in between. First, the superconducting lead (in blue), 20 nm thick Al, was formed. Next, the aluminum surface was oxidised to obtain tunnel barriers for the splitter. After that, the normal-state islands of 25 nm thick copper (in orange) were deposited, completing the splitter structure. As a third layer, a second 30 nm thick aluminum film was deposited followed by oxidation to produce the tunnel barriers for the detectors. The fabrication was then completed with the deposition of the 80 nm thick copper leads of the detectors. The four layer processing allows for independent tuning of the barrier transparencies for the splitter and the detectors. Fabricating the splitter first ensured that the aluminum reservoir is not directly connected to any of the extra non-operational shadow parts (seen e.g.~in the insets showing the detectors), thus preventing any proximity effect. Also, the normal-state islands of the splitter are designed so that the second aluminum layer does not overlap with them until several micrometers away from the tunnel junctions. The detectors use the inverse proximity effect to suppress the superconductivity of the small aluminum patches connected directly to the first, 25 nm thick Cu layer to obtain nearly normal-metallic charge detectors.

The tunnel junctions in the Cooper pair splitter were made with a distance of $l = 100\ \mathrm{nm}$. This distance was chosen to be shorter than the coherence length $\xi = 200\ \mathrm{nm}$ of the superconducting aluminum, hence allowing for a finite rate of Cooper pair splitting, which otherwise would be exponentially suppressed as $\exp(-l/\xi)$. By contrast, single-electron tunneling between the superconductor and the islands is suppressed as $\exp(-\Delta/k_BT)$, where $\Delta = \SI{200}{\micro eV}$ is the superconducting gap and $T = 50\, \mathrm{mK}\ll \Delta/k_B$ is the electronic temperature. Two-electron processes were thus the dominant charge transfer mechanism between the superconductor and the islands.  The filtering with the ground plane was paramount for obtaining the suppression as stray radiation causes excess single-electron tunneling. We note that measurements with the superconducting electrode in the normal-state were not performed, since the superconducting gap suppresses the sequential tunneling rates, which would otherwise exceed the detector bandwidth.

{\bf Experiments.} All experimental results presented here were obtained at the base temperature of a dry dilution refrigerator with an electronic temperature of 50 mK. The detectors were biased with $\SI{200}{\micro V}$, and the currents were measured with a digitizer after amplifying the signals with standard room temperature current preamplifiers with 1 kHz bandwidth. Voltages were applied to the gates to tune the charge detectors to a charge sensitive operation point and to tune the populations of the normal-state islands so that the two lowest-lying states in Fig.~\ref{fig2}\textbf{a} had equal probabilities. Time traces of the two detector currents were simultaneously recorded with a multichannel analog-to-digital converter at a sampling rate of 20 kHz, while adjusting the gate voltages in-between to compensate for slow drifts in the detector operation point or in the occupations of the two lowest charge states. The adjustment was done in a feedback loop by measuring a 60 s long time trace and extracting population probabilities from it. The electronics and measurement configuration were made identical on both detector sides to minimize timing differences.

{\bf Data analysis.} The 60 s long time traces were digitally filtered through a low-pass filter with a cut-off frequency of 200 Hz which sets the detector rise times to about 4 ms. All the results presented in our paper were obtained by analysing these filtered data. Time traces with the populations at the two lowest-lying charge states deviating considerably from each other or with the detector currents drifting so that it became difficult to identify charge states in the detector signal were excluded. The exact procedure and criteria are given in Supplementary Note 1. We then identified all tunneling events and the instances when they happened. These instances directly yield the experimental correlation functions in Figs.~\ref{fig1}\textbf{b} and \textbf{c} and the waiting time distributions in Figs.~\ref{fig3}\textbf{a} and \textbf{b}. In addition to the instances the theory curves require the detector broadening as input parameter which was obtained from the current noise of the detectors and their slew rate.

{\bf Theory.} The system is described by a rate equation, $\frac{d}{dt}|p(t)\rangle\!\rangle = \mathbf{L} |p(t)\rangle\!\rangle,$ where $|p(t)\rangle\!\rangle$ is a column vector containing the probabilities of being in the different charge states of the islands, and the rate matrix $\mathbf{L}$ contains the tunneling rates. The off-diagonal elements of $\mathbf{L}$ are given by the transition rates between the different charge states, while the diagonal elements contain the total escape rate (with a minus sign) from each charge state. In addition, we introduce a jump operator, $\mathbf{J}_\alpha$, describing transitions of type $\alpha$. The $g^{(2)}$-function for processes of type $\alpha$ and $\beta$ can then be obtained as
\begin{eqnarray}
\nonumber
g^{(2)}(\tau) &=& \frac{\langle \! \langle \mathbf{J}_\beta e^{\mathbf{L}\tau}\mathbf{J}_\alpha \rangle \! \rangle}{\langle \! \langle \mathbf{J}_\beta \rangle\! \rangle \langle\!\langle \mathbf{J}_\alpha \rangle\!\rangle }\theta(\tau) +  \frac{\langle \! \langle \mathbf{J}_\alpha e^{-\mathbf{L}\tau}\mathbf{J}_\beta \rangle \! \rangle}{\langle \! \langle \mathbf{J}_\beta \rangle\! \rangle \langle\!\langle \mathbf{J}_\alpha \rangle\!\rangle }\theta(-\tau)\\
&&+\frac{\langle \! \langle \mathbf{J}_{\alpha \beta}\rangle \! \rangle}{\langle \! \langle \mathbf{J}_\beta \rangle\! \rangle \langle\!\langle \mathbf{J}_\alpha \rangle\!\rangle }\delta(\tau),
\label{g2 def}
\end{eqnarray}
where $\langle \!\langle \mathbf{A}\rangle\!\rangle \equiv\langle \!\langle \tilde 0 | \mathbf{A} |p_s\rangle\!\rangle $ denotes the expectation value with respect to the steady-state fulfilling $\mathbf{L}|p_s\rangle\!\rangle =0$,  the vector representation of the trace operation is denoted as $\langle\! \langle \tilde 0|$, and $\theta(\tau)$ is the Heaviside step function. The $g^{(2)}$-function yields the normalized joint probability of detecting a transition of type $\alpha$ at some time and a transition of type $\beta$ at a time $\tau$ later. The last term in Equation~\eqref{g2 def} accounts for the instantaneous correlations between electrons belonging to the same two-electron process, described by a jump operator $ \mathbf{J}_{\alpha\beta}$. These correlations are convolved with a Gaussian distribution to take into account the timing jitter of the detectors.  The $g^{(2)}$-function approaches unity on timescales over which the correlations vanish.

The waiting time distribution can be obtained as
\begin{equation}
\mathcal{W}(\tau)= \frac{\langle\!\langle \mathbf{J}_\beta e^{(\mathbf{L}-\mathbf{J}_\beta)\tau} (\mathbf{J}_\alpha-\mathbf{J}_{\alpha \beta}) \rangle \!\rangle}{\langle \! \langle \mathbf{J}_\alpha \rangle \!\rangle }+\eta_0\delta(\tau),
\label{WTD def}
\end{equation}
with $\eta_0 \equiv \frac{\langle\!\langle \mathbf{J}_{\alpha \beta} \rangle \!\rangle}{\langle \! \langle \mathbf{J}_\alpha \rangle \!\rangle }$, which, just as for the $g^{(2)}$-function, includes the instantaneous correlations of two-electron processes. The waiting time distribution yields the probability density to observe a waiting time $\tau$ from a transition of type $\alpha$ has occurred until the first subsequent transition of type $\beta$ takes place. Similar to the $g^{(2)}$ function, we use a convolution with a Gaussian distribution to describe the timing jitter of the detectors. As a probability distribution, the waiting time distribution is normalized such that $\int_0^\infty d\tau \mathcal{W}(\tau) = 1$.

{\bf  Data availability.} The data that support the findings of this study are available from the corresponding authors upon reasonable request.

{\bf  Code availability.} The code to analyse the experimental data and to compute the theoretical results is available from the corresponding authors upon reasonable request.

\bibliography{main}

\vspace{0.5 cm}
{\bf Acknowledgements.} We thank J.~P.~Pekola and P.~Samuelsson for fruitful discussions. The work was supported financially by the QuantERA project “2D hybrid materials as a platform for topological quantum computing”, Swedish National Science Foundation, NanoLund and the Academy of Finland (project numbers 308515 and 331737). F.B. acknowledges support from  the European Union’s Horizon 2020 research and innovation programme under the Marie Sk\l odowska-Curie grant agreement number 892956. F.B., E.T.M., and C.F.~acknowledge the support by the Academy of Finland through the Finnish Centre of Excellence in Quantum  Technology (project numbers 312057 and 312299). We acknowledge the provision of facilities by Aalto University at OtaNano - Micronova Nanofabrication Centre.

{\bf Author contributions.} The experiment was carried out by A.R.~and V.F.M. The devices were fabricated by E.T.M. The theory was developed by F.B.~and C.F. All authors contributed to the discussion and analysis of the results and the writing of the manuscript. 

{\bf  Competing interests.} The authors declare no competing interests.

\end{document}


\title{Supplementary information: \\ Real-time observation of Cooper pair splitting showing strong non-local correlations}

\author{Antti Ranni}
\affiliation{NanoLund and Solid State Physics, Lund University, Box 118, 22100 Lund, Sweden}
\author{Fredrik Brange}
\author{Elsa T. Mannila}
\author{Christian Flindt}
\affiliation{Department of Applied Physics, Aalto University, 00076 Aalto, Finland}
\author{Ville F. Maisi}
\affiliation{NanoLund and Solid State Physics, Lund University, Box 118, 22100 Lund, Sweden}

\date{\today}

\maketitle

\section{Supplementary Note 1: Experimental details}

\subsection{A. Island occupations and device parameter values}

Strong Coulomb interactions allow us to control the occupation probabilities of the islands. For each island, the electrostatic energy reads $E_\alpha(n_\alpha)=E_{C\alpha}(n_\alpha-n_{G\alpha})^2$, where $n_\alpha$ is the number of excess electrons on the left or right island, $\alpha=L,R$. The charging energy is given as $E_{C\alpha}=e^2/2C_\alpha$ in terms of the total capacitance $C_\alpha$ of each island. We control the occupations via the parameters $n_{G\alpha}=C_{G\alpha}V_{G\alpha}/e$ by the gate voltages $V_{G\alpha}$ applied to the electrodes visible in Fig.~1\textbf{a} of the main text. By tuning the gate voltages so that $n_{G\alpha}=\nicefrac{1}{2}$, the charge states $n_\alpha=0$ and $n_\alpha=1$ become energetically degenerate with the corresponding charging energy diagram shown in Supplementary Fig.~\ref{sfig1} below. This choice makes the Cooper pair splitting process occur at no energy cost. The charging energies of the islands arise predominantly from the self-capacitance of the $\SI{12}{\micro m}$ long islands~\cite{maisi2011}, and we estimate the charging energies to be $E_{CL} = E_{CR} \sim \SI{40}{\micro eV}$ based on the values of Ref.~\citealp{maisi2011}. The tunnel junction transparency is estimated to be $\SI{170}{k\ohm \micro m^2}$ based on the average transparency of three reference tunnel junctions fabricated during the same fabrication round with resistances $\SI{30}{M\ohm}$, $\SI{42}{M\ohm}$ and $\SI{23}{M\ohm}$, and areas $\SI{75}{nm} \times \SI{68}{nm}$, $\SI{80}{nm} \times \SI{80}{nm}$ and $\SI{63}{nm} \times \SI{63}{nm}$ correspondingly.\\ 

Coulomb interactions between the metallic islands would favor elastic cotunneling over Cooper pair splitting as the splitting process would require extra energy for occupying both islands with an electron. To avoid this, the device is designed so that the islands are as far from each other as possible, and each island is much closer to the gate lines and the ground plane than the other island. At the junctions, where the islands are closest, the grounded superconductor screens the island-to-island Coulomb interactions. In the measured time traces, we observe that the charge state of one island does not considerably affect the state of the other. Hence, we conclude that the inter-dot charging energy is negligible, and the total energy of the two islands is simply given by the individual contributions from each island.

\begin{figure*}[b]
	\centering
	\includegraphics[width=0.48\textwidth]{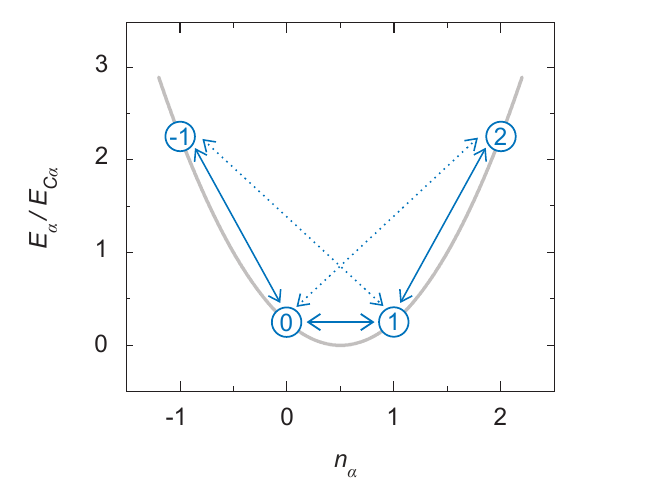}
	\caption{\textbf{Charging energy diagram.}
	Lowest-lying charge states of island $\alpha$, when the gate offset charge is tuned to $n_{g_\alpha}=\nicefrac{1}{2}$. With both islands at this degeneracy point, Cooper pair splitting is favorable, causing transitions from the charge state 0 to the charge state 1 in both islands. \label{sfig1}
		}
\end{figure*}

\subsection{B. Identification of charge states and tunneling events}

In the following, we explain how to determine the instantaneous charge states of the islands from the time traces. We also give the criteria by which some of the measured time traces were excluded from the analysis. The measured time traces, recorded at a sampling rate of 20 kHz simultaneously, were digitally filtered through a low-pass filter with a cut-off frequency of 200 Hz. This sets the detector rise times to $t_\mathrm{rise}=$ 4 ms. 60 s long time traces were then analysed one by one. The detector current (see Supplementary Fig.~\ref{sfig2}\textbf{a}) is divided into a histogram (Supplementary Fig.~\ref{sfig2}\textbf{b}) consisting of 100 bins. The two largest peaks in the histogram are identified as the current levels of the two energetically lowest charge states, namely $n_\alpha = 0$ and $n_\alpha = 1$, corresponding to the most common states of the islands. The peaks are found by first locating the global maximum $I_{\mathrm{h}}$ in the current histogram and removing all values around it within $\pm 2\sigma$, where $\sigma=2.5$ pA is the average standard deviation of the detector current at the charge states. From the remaining data in the histogram, the second maximum is located, and together these two maxima yield $I_{0}$ and $I_{1}$, such that $I_{0}$ is the state at the higher detector current, since  an increasing number of electrons decreases the detector current.\\

\begin{figure*}[t]
	\centering
	\includegraphics[width=0.96\textwidth]{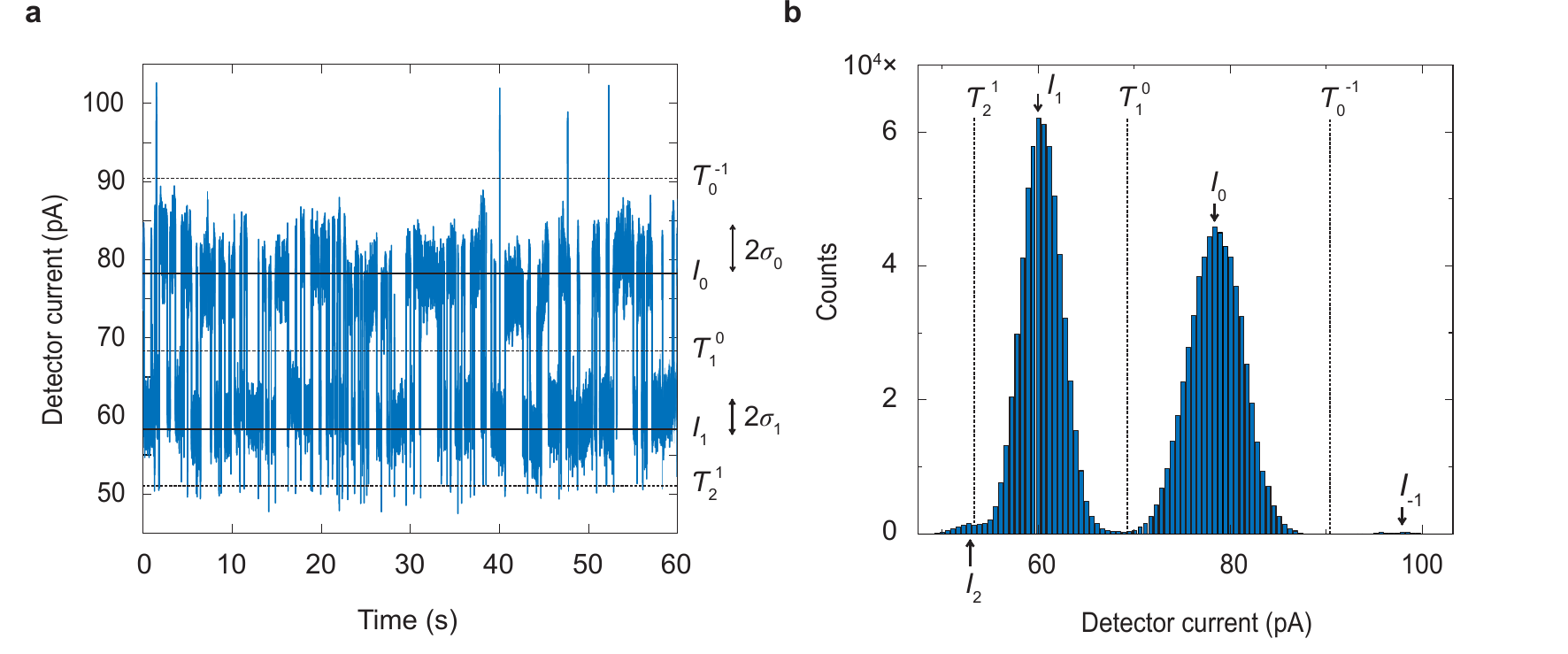}
	\caption{\textbf{Identification of charge states.}
	\textbf{a,} Example of a time trace, where the thresholds $\mathcal{T}_{j}^{i}$ (dashed lines) divide the current according to the different charge states. The current levels, $I_{0}$ and $I_{1}$, of the lowest-lying charge states are indicated with solid black lines. \textbf{b,} Detector current from \textbf{a} divided into 100 bins. The peaks in the histogram correspond to the current levels of the four charge states of the each island. The current in each level $j$ is given by a Gaussian distribution with standard deviation $\sigma_{j}$ as indicated in panel \textbf{a}. \label{sfig2}
		}
\end{figure*}

The ratio $N_{0}/N_{1}$ of the number of counts $N_i$ in the bins $I_{i}$ yields the relative occupations of the states $0$ and $1$. If the offset charge $n_{g_\alpha}$ on island $\alpha$ differs considerably from the degeneracy point, $n_{g_\alpha} = \nicefrac{1}{2}$, the time spent in the charge states 0 and 1 are not equal. We maintain the system at the degeneracy value of $N_{0}/N_{1} = 1$ by adjusting the gate voltages $V_{g_\alpha}$ in a feedback loop after the measurement of every time trace. To obtain the data from the main text, we use the time traces within the window $N_{0}/N_{1} = 1/2$ to $N_{0}/N_{1} = 2/1$.\\

Occasionally, the detector currents drift within the 60 s time trace such that one of the charge states drifts to the valley between $I_{0}$ and $I_{1}$, making it ambiguous to say whether an island is in the charge state 0 or 1. We discard these time traces using the following criteria: We examine how long time the detector current resides between $I_{0}$ and $I_{1}$ relative to the time spent at these two current levels by comparing $N_0$ and $N_1$ to the number of counts in between $N_\mathrm{valley}$. If there is a local maximum of the number of counts in the valley between the charge states 0 and 1 in the interval $[I_{1}+2\sigma,I_{0}-2\sigma]$ we take that as $N_{\mathrm{valley}}$. If there is no local maximum, we take the minimum in the same interval to be $N_{\mathrm{valley}}$. Then, the time trace is dismissed, if $N_{\mathrm{valley}}/\mathrm{min}(N_{0},N_{1}) > 0.05$. With this condition, the charge states stay on the correct side of the threshold levels $T_{i+1}^i$ of Supplementary Fig.~\ref{sfig2}\textbf{b} essentially at all times.\\  

To determine the charge state of the islands, we set a threshold $T_{1}^0$ halfway between the peaks $I_0$ and $I_1$. Similarly, if there is a local maximum in the histogram below $I_{1}-2\sigma$ (above $I_{0}+2\sigma$), we denote it as $I_{2}$ ($I_{-1}$) and set a threshold at the minimum between the peaks to distinguish the states $2$ and $-1$. If we do not observe the maxima $I_{-1}$ or $I_{2}$, we set no threshold and take all the data to be on state 0 or 1, correspondingly. Also, in order to avoid current noise induced false transitions, we have used a requirement that the detector current needs to get within $2\sigma_{i}$ of the charge state current $I_{i}$ ($i=0,1$) to register an event between $i$ and any of its neighbouring states. With the thresholds, we obtain the current ranges that allow us to determine the instantaneous charge state of the system and pinpoint the tunneling events as the points where at least one of the islands changes charge state.\\

\subsection{C. Measurements of the correlation functions}

The $g^{(2)}(\tau)$ correlation function describes how likely it is to observe an event at the time $\tau$ after another event took place. We determine the $g^{(2)}(\tau)$-functions in Fig. 1\textbf{b} and \textbf{c} of the main article directly based on the definition of the correlation function: We counted the number of events in a short time interval $\Delta \tau$ after time $\tau$ since the first event took place. The counting is straightforward after the identification of the tunneling events. The correlation function at time $\tau$ is determined then by normalizing the obtained counts with the appropriate normalization constant that depends on the total number of events, total measurement time and the width of the time interval as described in detail below. This normalization ensures the correct long time result of $g^{(2)}(\tau \rightarrow \pm\infty) = 1$. Importantly, the approach used here is the same for the auto-correlations and the cross-correlations with the only difference that in the auto-correlations, the two events are determined from the same detector and for the cross-correlations, the two events are from different detectors.\\


For the auto-correlation function $g^{(2)}(\tau)$ of Fig.~1\textbf{b} in the main text, we choose the counted events as follows: For each $0\rightarrow1$ event on the right island (the first event type), we count the number of $0\rightarrow1$ events taking place on the same island (the second event type) after a time $\tau$ around the time interval of $\Delta \tau = \SI{100}{ms}$. This yields the correlation function $g^{(2)}(\tau)$ at the time separation $\tau$, when divided by the normalization factor $(\dot{N}_\mathrm{avg}^{R})^2 \times t_\mathrm{tot} \times \Delta\tau$, where $\dot{N}_{\mathrm{avg}}^{R}$ is the total number of $0\rightarrow 1$ events on the right island per total measurement time $t_\mathrm{tot}$. The local one and two-electron tunneling processes were distinguished from each other before determining the correlation functions. The identification was made by taking two consecutive events happening within $t_\mathrm{rise}$ in the same detector to be local two-electron events and the rest to be sequential one-electron events. For example a local Andreev tunneling $-1\rightarrow1$ (consisting of closely happening consecutive $-1\rightarrow0$ and $0\rightarrow1$ transitions) was identified and did not contribute to $0\rightarrow1$ events in correlation functions. \\

In the auto-correlation measurements, the detector rise time $t_\mathrm{rise}$ limits the smallest $\tau$ that one may observe since the detector needs to respond before it can detect the next event. The time resolution of the auto-correlation measurement is visualized in Supplementary Fig.~\ref{sfig3}\textbf{a}, where we present a typical time trace yielding one count to the auto-correlation function. The count is obtained as two $0\rightarrow1$ events indicated with vertical dashed lines are observed in the right detector. The time $\tau$ indicates the time separation between the events and $t_\mathrm{rise}$ the detector rise time. This limitation is illustrated in Supplementary Fig.~\ref{sfig4} that shows the auto-correlation function for sequential tunneling processes $0\leftrightarrow1$ on the right island. It is obtained similarly to the auto-correlation function in Fig.~1\textbf{b} of the main article with the only difference that here we account for both single-electron tunneling into and out of the island. An ideal detector with infinitely fast rise time would result in a flat $g^{(2)}(\tau)$ function because the considered tunneling events are uncorrelated, having the value one at each point in time even at $\tau=0$ like the solid theory curve. Since our detectors have a finite rise time set by the bandwidth, we see a dip in the $g^{(2)}(\tau)$ function around zero time. The width of the dip matches $t_\mathrm{rise}$, hence demonstrating the dead time of the detector limiting the short detection times for the auto-correlation function.\\

Returning now to the auto-correlation data of the $0\rightarrow1$ events presented in Fig.~1\textbf{b} of the main article, another timescale sets in: When detecting the two $0\rightarrow1$ events, the system needs to switch back to the initial state $0$ before the second $0\rightarrow1$ event can be detected. Therefore the tunneling timescale to return from $1$ to $0$ yields a dip to the auto-correlation function. As seen in Fig.~1\textbf{b} of the main article, this dip is roughly one second wide. Thus, the much shorter $t_\mathrm{rise}$ does not have a considerable effect on our results on auto-correlation presented in the main text.\\  

\begin{figure*}[t]
	\centering
	\includegraphics[width=\textwidth]{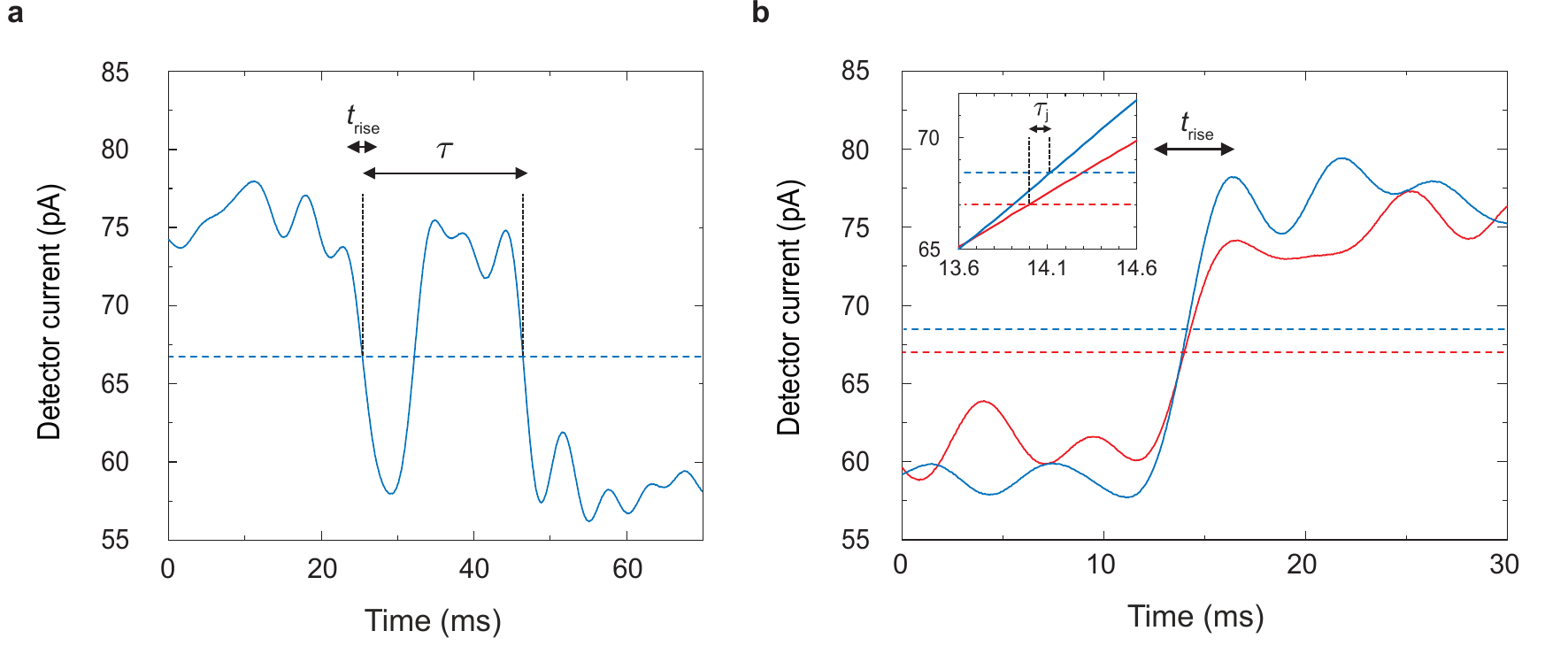}
	\caption{\textbf{Time scales for detecting the tunneling events.}
    	\textbf{a,} A typical time trace of the right detector yielding one count to the auto-correlation measurement. The solid line is the measured detector signal and the horizontal dashed line indicates the threshold between charge states 0 (above the line) and 1 (below). The time $\tau$ is the separation between two $0\rightarrow1$ transitions. The detector rise time $t_\mathrm{rise}$ sets a limit to how quickly the two events can be observed with the same detector. \textbf{b,} A typical time trace yielding one count to the cross-correlation measurement. Detector currents for the left and the right islands are shown in red and blue curves respectively. The detectors switch almost simultaneously from the charge state 1 to 0. The observed time separation $\tau_{\mathrm{j}} \ll t_\mathrm{rise}$ between the $1\rightarrow0$ events on the two distinct islands is indicated in the zoom-in of the inset. \label{sfig3}
		}
\end{figure*}

\begin{figure*}[t]
	\centering
	\includegraphics[width=0.48\textwidth]{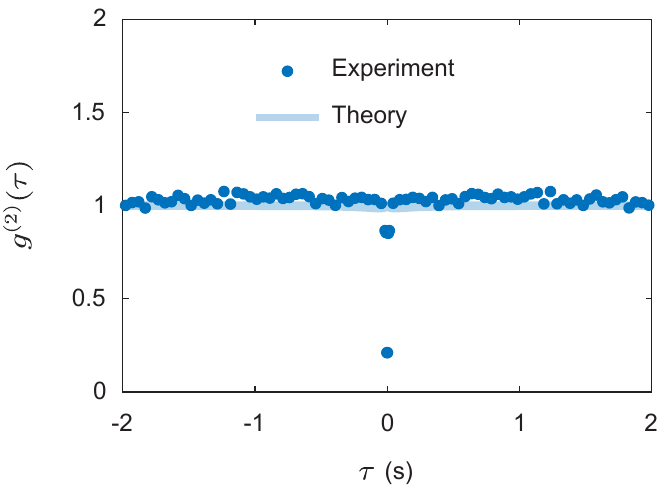}
	\caption{\textbf{Influence of the finite detector rise time to correlation measurements.}
	Auto-correlation function for $0\rightarrow1$ and $1\rightarrow0$ processes on the right island. The experimental points close to zero time show a millisecond-wide dip due to the finite rise time of the detector. \label{sfig4}
		}
\end{figure*}

For the cross-correlation function $g^{(2)}_x(\tau)$ of Fig.~1\textbf{c} in the main text, we choose the counted events on separate islands. For the first event type we choose $0\leftrightarrow1$ transitions on the left island and count the number of $0\leftrightarrow1$ transitions on the right island after a time $\tau$ around the short time interval of $\Delta \tau = \SI{150}{\micro s}$. Detector currents are recorded every $\SI{50}{\micro s}$ and the interval is chosen commensurate to this such that each interval contains three sampling points. This is achieved by centering one interval $\Delta \tau$ at $\tau=0$ and adding intervals next to each other towards positive and negative $\tau$ values. The normalization factor for the cross-correlation is $\dot{N}_{\mathrm{avg}}^{L}\times \dot{N}_{\mathrm{avg}}^{R}\times t_\mathrm{tot}\times \Delta\tau$ as we now account for the total number of events $\dot{N}_{\mathrm{avg}}^{\alpha}$ between charge states 0 and 1 on both islands $\alpha$ per total measurement time $t_\mathrm{tot}$.\\


Supplementary Figure~\ref{sfig3}\textbf{b} presents the timing resolution of the cross-correlation measurements. The measured time traces exhibit nearly simultaneously switching that yields one count to the cross-correlation function of Fig.~1\textbf{c} of the main article. The time difference $\tau$ of the splitting events arises from the detector timing jitter $\tau_\mathrm{j}$ presented in the inset. Interestingly, the restriction on the time resolution from the finite rise time $t_\mathrm{rise}$ is lifted when monitoring events with two distinct detectors. When having an own detector for both tunneling events, neither of the detectors needs to recover from detection of the first event and hence the detector dead time is not relevant. We observe this directly in the measured time traces of Supplementary Fig.~\ref{sfig3}\textbf{b}. The observed time separation $\tau_\mathrm{j}$ between the two transitions is much smaller than the rise time $t_\mathrm{rise}$ of the individual detectors. This finding is valid for all the events at the correlation peak of Fig.~1\textbf{c} of the main article. Hence, we demonstrate a sub-millisecond time resolution despite the detector rise time is in the millisecond range.\\

The relevant limiting factor for the time resolution of the cross-correlation measurement arises from the detector noise. Noise in the measured electrical current translates into noise in the timing of the detectors~\cite{Maichen2006}. This is known as the timing jitter that we denoted above with $\tau_\mathrm{j}$. The relation of the current noise $\sigma_{I_{i}}$ and the timing jitter (i.e. the 'time noise') $\sigma_{\tau_\mathrm{j}}$ is given by the slew rate $dI/dt$ of the detector. The slew rate characterizes how quickly the detector responds to a change and is directly obtained from the derivatives of the detector response of Supplementary Fig.~\ref{sfig3}\textbf{b} near the thresholds. Similarly the current noise $\sigma_{I_{i}}$ is obtained from the standard deviation of the current within the charge state $i$, cf. Supplementary Fig.~\ref{sfig2}. With these, we obtain
$\sigma_{\tau_{\mathrm{j}}(i\rightarrow f)} = |dI/dt|^{-1} \sigma_{I_{i}}$. The slew rates, obtained directly from the traces, are 6.7 pA/ms and 8.4 pA/ms for $0\rightarrow1$ and $1\rightarrow0$ transitions on the right island respectively. On the left island the slew rate is 5.4 pA/ms to both directions. The current noises for the left detector are $\sigma_{I_{0}}=2.2$ pA, $\sigma_{I_{1}}=1.7$ pA, and for the right detector $\sigma_{I_{0}}=2.8$ pA and $\sigma_{I_{1}}=2.2$ pA. The slew rates together with the current noises yield the detector timing jitters  $\sigma_{\tau_{\mathrm{j}}(0\rightarrow 1)}=410$ $\mu$s, $\sigma_{\tau_{\mathrm{j}}(1\rightarrow 0)}=310$ $\mu$s for the left detector and $\sigma_{\tau_{\mathrm{j}}(0\rightarrow 1)}=420$ $\mu$s and $\sigma_{\tau_{\mathrm{j}}(1\rightarrow 0)}=260$ $\mu$s for the right detector. Then the detector broadening of the $g_{x}^{(2)}(\tau)$ function is determined as weighted sums of each participating charge state, see Equation~(\ref{eq:broad1}). The timing jitters together with the tunneling rates below are all the input parameters needed for the theory curves without any further fitting.


\subsection{D. Tunneling rates}

\begin{figure*}[t]
	\centering
	\includegraphics[width=0.98\textwidth]{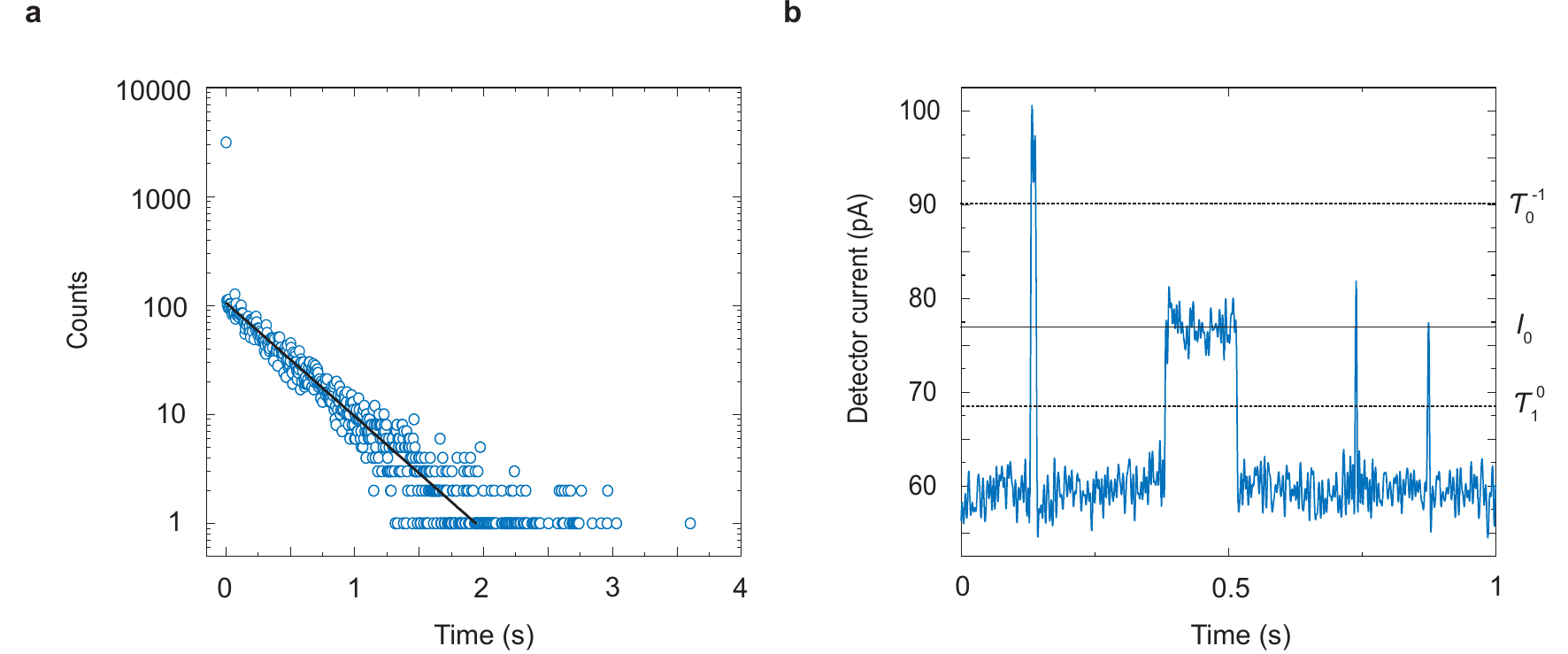}
    \caption{\textbf{Life-time distribution.}
	\textbf{a,} Time spent in the charge state 0 of the right island before a tunneling event out of it. The bin size is 4 ms, and the first point arises due to Andreev tunneling that started from another state. The long-time statistics follow an exponential distribution which determines the life-time of the state. \textbf{b,} The leftmost peak shows a local Andreev tunneling from $n_R = 1\rightarrow-1$ and its time reversal process to opposite direction. Duration the detector current spends at $n_R=0$ is less than the detector rise time of 4 ms. The wide peak in the middle of the graph corresponds to a sequential tunneling $n_R=1\rightarrow0$ and its time reversal process. The two sharp peaks on the right side are local Andreev tunneling events which take the right island $n_R=1\rightarrow-1$ and are followed by another local Andreev back to $n_R=1$ but these events happen in a time shorter than the detector rise time and thus it appears as if there were tunneling sequences $n_R=1\rightarrow0\rightarrow1$. \label{sfig5}
	}
\end{figure*}

To identify which tunneling process took place with each event we followed the procedure of Ref.~\citealp{maisi2011}: We interpret all events that take place within the detector rise time of 4 ms at the same island to belong to the same event. Supplementary Figure~\ref{sfig5} presents typical cases taking place around the charge state $n_R = 0$ of the right island. Panel a summarizes the life-time distribution of the charge state if all events at the detector would be treated as sequential single-electron events. In this case, the bin at the shortest time within the detector rise time have anomalously many events whereas all the rest of the data follows an exponential distribution as expected for tunneling starting from $n_R = 0$. The reason for the anomalously high first data point is that those events belong to a process that started from another charge state and should not be counted into $n_R = 0$ state~\cite{maisi2011}. Panel b shows an example time trace with such events.\\

In Supplementary Fig.~\ref{sfig5}\textbf{b} we see in the beginning of the time trace a transition $n_R = 1 \rightarrow 0$ followed immediately within the detector rise time by $n_R = 0 \rightarrow -1$ yielding us a two electron local Andreev tunneling event. The other case contributing to short time counts of panel a is also shown in Supplementary Fig.~\ref{sfig5}\textbf{b}. In this case the systems seems to take transition $1 \rightarrow 0$ followed quickly by another event with $0 \rightarrow 1$. Two such cases are visible in the end of the time trace. However, based on the lifetime distribution, there are too many of the latter return transitions $0 \rightarrow 1$ for them to arise from the $0 \rightarrow 1$ single-electron transition. Therefore, we interpret this course of events to arise from two local Andreev transitions, one with $n_R = 1 \rightarrow -1$ followed by $n_R = -1 \rightarrow 1$. The second transition takes place quickly after the first one as it has an energy gain, see the energy diagram of Supplementary Fig.~\ref{sfig1}. The tunneling rate for the return process is so fast that the detector does not have time to reach the final state before starting to return to the initial state. Thus the $1 \leftrightarrow -1$ transition appears as if it would be between $1 \leftrightarrow 0$. Due to this ambiguity, we exclude the cases with a fast return event (lifetime at state 0 shorter than 4 ms) from the determination of the tunneling rates. We note that these events take place locally on one of the islands and hence they do not impact the results presented in the main article. If we take them into the analysis as Andreev events the correlation function results remain unchanged and the hold time would still remain two orders of magnitude longer than the detection time.\\

With the above protocol we distinguish different local tunneling processes. In addition, we identify the crossed processes as those $0\leftrightarrow1$ events that take place on different islands within the 1.5 ms time window obtained from the cross-correlation function and finally determine the tunneling rates by counting how many times the corresponding tunneling process occurred and divide with the time spent in the initial state.
The tunneling rates for local processes are listed in Supplementary Table~\ref{table1}.  The non-local crossed tunneling rates are $\Gamma_{\mathrm{L=0\rightarrow 1}}^{\mathrm{R=0\rightarrow 1}}=14$ mHz for Cooper pair splitting, $\Gamma_{\mathrm{L=1\rightarrow 0}}^{\mathrm{R=1\rightarrow 0}}=140$ mHz for Cooper pair assembling, $\Gamma_{\mathrm{L=1\rightarrow 0}}^{\mathrm{R=0\rightarrow 1}}=25$ mHz for elastic cotunneling from the left island into the right one, and $\Gamma_{\mathrm{L=0\rightarrow 1}}^{\mathrm{R=1\rightarrow 0}}=96$ mHz for elastic cotunneling from the right island into the left one.

\begin{table}[t]
            \caption{Tunneling rates for the processes on individual islands (L for the left island and R for the right one). \label{table1}} \vspace{3pt}
				\begin{tabular}{ccccccccccc}
				    \multicolumn{1}{r}{}
					& \multicolumn{1}{l}{$\Gamma_{1\rightarrow 0}$ (Hz)}
					& \multicolumn{1}{l}{$\Gamma_{0\rightarrow 1}$ (Hz)}
					& \multicolumn{1}{l}{$\Gamma_{2\rightarrow 1}$ (Hz)}
					& \multicolumn{1}{l}{$\Gamma_{1\rightarrow 2}$ (Hz)}
					& \multicolumn{1}{l}{$\Gamma_{0\rightarrow -1}$ (Hz)}
					& \multicolumn{1}{l}{$\Gamma_{-1\rightarrow 0}$ (Hz)}
					& \multicolumn{1}{l}{$\Gamma_{2\rightarrow 0}$ (Hz)}
					& \multicolumn{1}{l}{$\Gamma_{0\rightarrow 2}$ (Hz)}
					& \multicolumn{1}{l}{$\Gamma_{1\rightarrow -1}$ (Hz)}
					& \multicolumn{1}{l}{$\Gamma_{-1\rightarrow 1}$ (Hz)} \\ \hline
					L\phantom{L} & 0.11 & 0.084 & 700 & 0.017 & 0.21 & 7.2 & 160 & 0.0026 & 0.11 & 3.8 \\
					R\phantom{L} & 1.6 & 2.3 & 9.2 & 0.69 & 0.024 & 4.4 & 26 & 0.0080 & 0.32 & 29 \\ \hline
				\end{tabular}
		\end{table}

\subsection{E. Waiting time distributions}

The waiting time distribution in Fig.~3\textbf{a} of the main text is obtained by determining the time duration $\tau$ from a $1 \rightarrow 0$ transition to the next $1 \rightarrow 0$ transition locally on the right detector. These transitions are either sequential tunneling events solely on the right island or arise from crossed processes moving the right system from $n_R = 1$ to $n_R = 0$. Between two $1\rightarrow0$ transitions, any other type of tunneling process is allowed to take place. For example, after a $1\rightarrow0$ transition we can have a sequence $0\rightarrow1\rightarrow-1\rightarrow1$ before a single electron tunnels out again from the state 1. Hence, the waiting time describes the time scale one needs to wait until the same event occurs again. To obtain the distribution, we again count the number transitions taking a time $\tau$ since the previous transition around a short time interval of $\Delta \tau$. This yields the waiting time distribution $\mathcal{W}(\tau)$, when divided by the normalization factor $N \times \Delta \tau$ where $N$ is the total number of counts to all bins.\\

For the cross-waiting time distribution in Fig.~3\textbf{b} of the main text, we consider the simultaneously measured time traces on the two islands and find the waiting times $\tau$ from each $1\rightarrow0$ transition on the left island to the consecutive $1\rightarrow0$ event on the right island. During the waiting time $\tau$ any tunneling process on the left island is allowed, even another $1\rightarrow0$ transition. The cross-waiting time distribution $\mathcal{W}_{x}(\tau)$ is computed the same way as $\mathcal{W}(\tau)$.        

\section{Supplementary Note 2: Theoretical model}

In the following, we develop a theoretical model of the Cooper pair split, which we use to derive all of the theoretical results in the main text. To this end, we employ a rate equation to describe the dynamics of the two islands,
\begin{equation}
\frac{d}{dt}|p(t)\rangle\!\rangle = \mathbf{L} |p(t)\rangle\!\rangle,
\end{equation}
where $|p(t)\rangle\!\rangle = (p_1 \quad p_2 \quad ... \quad p_{16})^T$ is a column vector containing the probabilities $p_i$ of being in the 16 different charge states of the islands, and $\mathbf{L}$ is a rate matrix. $\mathbf{L}$ has the off-diagonal elements
$\mathbf{L}_{ij} = \Gamma_{j\rightarrow i}$, $i\neq j$, and the diagonal elements
$\mathbf{L}_{jj} = -\sum_{i\neq j} \Gamma_{j\rightarrow i}$, with $\Gamma_{j\rightarrow i}$ being the transition rate from the $j$th to the $i$th state. To be explicit, the off-diagonal part of $\mathbf{L}$ reads\\
\begin{eqnarray}
\nonumber
\begin{pmatrix}
0& \Gamma^R_{0\rightarrow \text{-}1}& \Gamma^R_{1\rightarrow \text{-}1} & 0 & \Gamma^L_{0\rightarrow \text{-}1}& 0 & 0 & 0 & \Gamma^L_{1\rightarrow \text{-}1} & 0 & 0 & 0 & 0 & 0 & 0 & 0 \\
\Gamma^R_{\text{-}1\rightarrow 0}& 0 & \Gamma^R_{1\rightarrow 0} & \Gamma^R_{2\rightarrow 0} & 0& \Gamma^L_{0\rightarrow \text{-}1} & 0 & 0 & 0 & \Gamma^L_{1\rightarrow \text{-}1} & 0 & 0 & 0 & 0 & 0 & 0 \\
\Gamma^R_{\text{-}1\rightarrow 1}& \Gamma^R_{0\rightarrow 1}& 0 & \Gamma^R_{2\rightarrow 1} & 0 & 0 & \Gamma^L_{0\rightarrow \text{-}1} & 0 & 0 & 0 & \Gamma^L_{1\rightarrow \text{-}1} & 0 & 0 & 0 & 0 & 0 \\
0 & \Gamma^R_{0\rightarrow 2}& \Gamma^R_{1\rightarrow 2} & 0 & 0 & 0 & 0 & \Gamma^L_{0\rightarrow \text{-}1} & 0 & 0 & 0 & \Gamma^L_{1\rightarrow \text{-}1} & 0 & 0 & 0 & 0 \\
\Gamma^L_{\text{-}1\rightarrow 0} & 0& 0 & 0 & 0 & \Gamma^R_{0\rightarrow \text{-}1} & \Gamma^R_{1\rightarrow \text{-}1}& 0 & \Gamma^L_{1\rightarrow 0} & 0 & 0 & 0 & \Gamma^L_{2\rightarrow 0} & 0 & 0 & 0 \\
0 & \Gamma^L_{\text{-}1\rightarrow 0}& 0 & 0 & \Gamma^R_{\text{-}1\rightarrow 0} & 0 & \Gamma^R_{1\rightarrow 0}& \Gamma^R_{2\rightarrow 0} & 0 & \Gamma^L_{1\rightarrow 0} & \Gamma^\text{CAR}_\text{in} & 0 & 0 & \Gamma^L_{2\rightarrow 0} & 0 & 0 \\
0 & 0 & \Gamma^L_{\text{-}1\rightarrow 0} & 0 & \Gamma^R_{\text{-}1\rightarrow 1} & \Gamma^R_{0\rightarrow 1} & 0 & \Gamma^R_{2\rightarrow 1} & 0 & \Gamma^\text{EC}_{L\rightarrow R} & \Gamma^L_{1\rightarrow 0} & 0 & 0 & 0 & \Gamma^L_{2\rightarrow 0} & 0 \\
0 & 0 & 0 & \Gamma^L_{\text{-}1\rightarrow 0} & 0 & \Gamma^R_{0\rightarrow 2} & \Gamma^R_{1\rightarrow 2} & 0 & 0 & 0 & 0 & \Gamma^L_{1\rightarrow 0} & 0 & 0 & 0 & \Gamma^L_{2\rightarrow 0} \\
\Gamma^L_{\text{-}1\rightarrow 1} & 0 & 0 & 0 & \Gamma^L_{0\rightarrow 1} & 0 & 0 & 0 & 0 & \Gamma^R_{0\rightarrow \text{-}1} & \Gamma^R_{1\rightarrow \text{-}1} & 0 & \Gamma^L_{2\rightarrow 1} & 0 & 0 & 0 \\
0 & \Gamma^L_{\text{-}1\rightarrow 1} & 0 & 0 & 0 & \Gamma^L_{0\rightarrow 1} & \Gamma^\text{EC}_{R\rightarrow L} & 0 & \Gamma^R_{\text{-}1\rightarrow 0} & 0 & \Gamma^R_{1\rightarrow 0} & \Gamma^R_{2\rightarrow 0} & 0 & \Gamma^L_{2\rightarrow 1} & 0 & 0 \\
0 & 0 & \Gamma^L_{\text{-}1\rightarrow 1} & 0 & 0 & \Gamma^\text{CAR}_\text{out} & \Gamma^L_{0\rightarrow 1} & 0 & \Gamma^R_{\text{-}1\rightarrow 1} & \Gamma^R_{0\rightarrow 1} & 0 & \Gamma^R_{2\rightarrow 1} & 0 & 0 & \Gamma^L_{2\rightarrow 1} & 0 \\
0 & 0 & 0 & \Gamma^L_{\text{-}1\rightarrow 1} & 0 & 0 & 0 & \Gamma^L_{0\rightarrow 1} & 0 & \Gamma^R_{0\rightarrow 2} & \Gamma^R_{1\rightarrow 2} & 0 & 0 & 0 & 0 & \Gamma^L_{2\rightarrow 1} \\
0 & 0 & 0 & 0 & \Gamma^L_{0\rightarrow 2} & 0 & 0 & 0 & \Gamma^L_{1\rightarrow 2} & 0 & 0 & 0 & 0 & \Gamma^R_{0\rightarrow \text{-}1} & \Gamma^R_{1\rightarrow \text{-}1} & 0 \\
0 & 0 & 0 & 0 & 0 & \Gamma^L_{0\rightarrow 2} & 0 & 0 & 0 & \Gamma^L_{1\rightarrow 2} & 0 & 0 & \Gamma^R_{\text{-}1\rightarrow 0} & 0 & \Gamma^R_{1\rightarrow 0} & \Gamma^R_{2\rightarrow 0} \\
0 & 0 & 0 & 0 & 0 & 0 & \Gamma^L_{0\rightarrow 2} & 0 & 0 & 0 & \Gamma^L_{1\rightarrow 2} & 0 & \Gamma^R_{\text{-}1\rightarrow 1} & \Gamma^R_{0\rightarrow 1} & 0 & \Gamma^R_{2\rightarrow 1} \\
0 & 0 & 0 & 0 & 0 & 0 & 0 & \Gamma^L_{0\rightarrow 2} & 0 & 0 & 0 & \Gamma^L_{1\rightarrow 2} & 0 & \Gamma^R_{0\rightarrow 2} & \Gamma^R_{1\rightarrow 2} & 0 \\ 
\end{pmatrix}.
\label{Off-diag L}
\end{eqnarray}\\
Together with the diagonal part, each column of the rate matrix sums to zero, ensuring that the total probability of being in any of the states is preserved over time. The jump operator $\mathbf{J}_\alpha$ of a certain kind of transition $\alpha$ contains all off-diagonal matrix elements of $\mathbf{L}$ corresponding to the processes involving that transition. We note that the crossed Andreev reflections enter here as one of the possible processes. For instance, the jump operator $\mathbf{J}_{1\rightarrow0}^L$ contains all off-diagonal matrix elements of $\mathbf{L}$ that involve the transition $1\rightarrow 0$ on the left island, including crossed Andreev and elastic cotunneling processes.

\subsection{A. Second-order correlation functions}
We define the $g^{(2)}$-correlation function of two types of transitions $\alpha$ and $\beta$ as \cite{PhysRevA.39.1200}
\begin{equation}
g^{(2)} (t,t+\tau) = \frac{P_{\beta \alpha}(t+\tau,t)}{P_\beta(t+\tau)P_\alpha(t)},
\label{g2 as probabilities}
\end{equation}
where $P_{\alpha}(t)$ is the probability density of transition $\alpha$ taking place at time $t$, and $P_{\beta \alpha}(t+\tau,t)$ is the joint probability density of transition $\alpha$ taking place at $t$ and transition $\beta$ at $t+\tau$. For a steady state, the $g^{(2)}$-function depends only on the time difference between the transitions, $g^{(2)} (t,t+\tau) \equiv g^{(2)} (\tau)$. In that case, we may express Eq.~\eqref{g2 as probabilities} in terms of jump operators as \cite{PhysRevB.85.165417}
\begin{equation}
g^{(2)}(\tau) = \frac{\langle \! \langle \mathbf{J}_\beta e^{\mathbf{L}\tau}\mathbf{J}_\alpha \rangle \! \rangle}{\langle \! \langle \mathbf{J}_\beta \rangle\! \rangle \langle\!\langle \mathbf{J}_\alpha \rangle\!\rangle }\theta(\tau) +  \frac{\langle \! \langle \mathbf{J}_\alpha e^{-\mathbf{L}\tau}\mathbf{J}_\beta \rangle \! \rangle}{\langle \! \langle \mathbf{J}_\beta \rangle\! \rangle \langle\!\langle \mathbf{J}_\alpha \rangle\!\rangle }\theta(-\tau)+\frac{\langle \! \langle \mathbf{J}_{\alpha \beta}\rangle \! \rangle}{\langle \! \langle \mathbf{J}_\beta \rangle\! \rangle \langle\!\langle \mathbf{J}_\alpha \rangle\!\rangle }\delta(\tau),
\label{g2 def}
\end{equation}
where $\langle \!\langle \mathbf{A}\rangle\!\rangle \equiv\langle \!\langle \tilde 0 | \mathbf{A} |p_s(t)\rangle\!\rangle $ is the expectation value with respect to the steady state fulfilling $\mathbf{L}|p_s(t)\rangle\!\rangle =0$, $\langle\! \langle \tilde 0|$ is the vector representation of the trace operation, $\theta(\tau)$ is the Heaviside step function, and $\delta(\tau)$ is the Dirac delta function. The last term in Equation~\eqref{g2 def} accounts for instantaneous correlations between electrons belonging to the same two-electron process. Such processes are described by two-electron superoperators $\mathbf{J}_{\alpha\beta} = (\mathbf{J}_{\alpha}\circ \mathbf{J}_{\beta})^{\circ1/2}(1-\delta_{\alpha \beta})$, with $\circ$ denoting Hadamard operations (element-wise operations) and $\delta_{\alpha \beta}$ the Kronecker delta function, containing all the matrix elements involving both transitions $\alpha$ and $\beta$. We note here that the instantaneous correlations are strongly affected by the timing jitter of the detectors, which effectively broadens the delta spike in time. The timing jitter is accounted for by convolving the correlation function with a Gaussian distribution of width $\sigma_D$, as discussed below for the cross-correlation $g^{(2)}_x(\tau)$-function.

\subsubsection{1. Equation (1) in the main text}
We first consider the auto-correlation $g^{(2)}$-function for electrons tunneling out of the right island into the superconductor. To this end, we set $\mathbf{J}_\alpha = \mathbf{J}_\beta =\mathbf{J}^R_{1\rightarrow 0}$ and $\mathbf{J}_{\alpha \beta} = 0$ in Equation~\eqref{g2 def}, yielding the expression
\begin{equation}
g^{(2)}(\tau) = \frac{\langle \! \langle \mathbf{J}^R_{1\rightarrow 0} e^{\mathbf{L}|\tau|}\mathbf{J}^R_{1\rightarrow 0} \rangle \! \rangle}{\langle \! \langle \mathbf{J}^R_{1\rightarrow 0} \rangle\! \rangle^2 }.
\label{Eq 1 in main text}
\end{equation}
Next, we make use of the fact that the two-electron processes involving also the left island are slow compared to the dynamics of the right island. To compute the auto-correlation $g^{(2)}$-function, we may therefore treat the right island as an approximately separate four-level system, with the rate matrix
\begin{equation}
	\mathbf{L}_R = \begin{pmatrix}
	-(\Gamma^R_{\text{-}1\rightarrow 0}+\Gamma^R_{\text{-}1\rightarrow 1}) & \Gamma^R_{0\rightarrow \text{-}1} & \Gamma^R_{1\rightarrow \text{-}1} & 0\\
	\Gamma^R_{\text{-}1 \rightarrow 0} & -(\Gamma^R_{0\rightarrow \text{-}1}+\Gamma^R_{0\rightarrow 1}+\Gamma^R_{0\rightarrow 2}) & \Gamma^R_{1\rightarrow 0} & \Gamma^R_{2\rightarrow 0} \\
	\Gamma^R_{\text{-}1 \rightarrow 1} & \Gamma^R_{0 \rightarrow 1} & -(\Gamma^R_{1\rightarrow \text{-}1}+\Gamma^R_{1\rightarrow 0}+\Gamma^R_{1\rightarrow 2}) & \Gamma^R_{2\rightarrow 1} \\
	0 & \Gamma^R_{0 \rightarrow 2} & \Gamma^R_{1 \rightarrow 2} &  -(\Gamma^R_{2\rightarrow 0}+\Gamma^R_{2\rightarrow 1})
	\end{pmatrix}.
	\label{L4}
\end{equation}
With the total rates out of the states $-1$ and $+2$ being much larger than all other rates, we find that the right island effectively behaves as a two-level system with the high-energy states $-1$ and $+2$ as short-lived virtual states, resulting in the correlation function
\begin{equation}
	g^{(2)}(\tau) = 1-\exp\left[-\left(\widetilde \Gamma_{0\rightarrow 1}^R+\widetilde \Gamma^R_{1\rightarrow 0}\right) |\tau|\right],
	\label{g2 auto}
\end{equation}
which is the expected functional form of the $g^{(2)}$-function of a two-level system. However, here $\widetilde \Gamma_{0\rightarrow 1}^R$ and $\widetilde \Gamma^R_{1\rightarrow 0}$ are the \emph{effective} transition rates between the two levels, which to leading order in the ratio between the total rates out of the virtual states and the other transition rates are
\begin{eqnarray}
\nonumber
\widetilde \Gamma^R_{0\rightarrow 1} &=& \Gamma^R_{0\rightarrow 1}+\frac{\Gamma^R_{0\rightarrow -1}\Gamma^R_{-1\rightarrow 1}}{\Gamma^R_{-1\rightarrow 0}+\Gamma^R_{-1\rightarrow 1}}+\frac{\Gamma^R_{0\rightarrow 2} \Gamma^R_{2\rightarrow 1}}{\Gamma^R_{2\rightarrow 0}+\Gamma^R_{2\rightarrow 1}} = 2.3 \text{ s}^{-1}, \\
\widetilde \Gamma^R_{1\rightarrow 0} &=& \Gamma^R_{1\rightarrow 0}+\frac{\Gamma^R_{1\rightarrow -1}\Gamma^R_{-1\rightarrow 0}}{\Gamma^R_{-1\rightarrow 0}+\Gamma^R_{-1\rightarrow 1}}+\frac{\Gamma^R_{1\rightarrow 2} \Gamma^R_{2\rightarrow 0}}{\Gamma^R_{2\rightarrow 0}+\Gamma^R_{2\rightarrow 1}} = 2.2 \text{ s}^{-1},
\label{Effective rates auto-g2}
\end{eqnarray}
where the last two terms in each equation describe the contributions from the virtual states in addition to the bare rates $\Gamma^R_{0\rightarrow 1}$ and $\Gamma^R_{1\rightarrow0}$. Introducing $\gamma \equiv \widetilde \Gamma_{0\rightarrow 1}^R+\widetilde \Gamma^R_{1\rightarrow 0} = 4.5 \text{ s}^{-1}$, we recover Equation~(1) of the main text from Equation~\eqref{g2 auto}. An exact numerical calculation based on Equation~\eqref{Eq 1 in main text} shows good agreement with the analytic result in Equation~\eqref{g2 auto}, although the exact numerical value of $\gamma$ is slightly higher due to the contributions from the two-particle processes involving both islands that are omitted in the approximations above.

\subsubsection{2. Equation (2) in the main text}
Next, we consider the cross-correlation $g^{(2)}$-function between electrons tunneling into or out of the left island and electrons tunneling into or out of the right island. In this case, we have $\mathbf{J}_\alpha = \mathbf{J}^{L}_{0\rightarrow 1}+\mathbf{J}^{L}_{1\rightarrow 0}$, $\mathbf{J}_\beta =\mathbf{J}^R_{0\rightarrow 1}+\mathbf{J}^R_{1\rightarrow 0}$ and $\mathbf{J}_{\alpha \beta} =\mathbf{J}^\text{CAR}_\text{in}+\mathbf{J}^\text{CAR}_\text{out}+\mathbf{J}^\text{EC}_{L\rightarrow R}+\mathbf{J}^\text{EC}_{R\rightarrow L}$ in Equation~\eqref{g2 def} and obtain
\begin{equation}
\begin{split}
g^{(2)}_x(\tau) &=\frac{\langle \! \langle (\mathbf{J}^{R}_{0\rightarrow 1}+\mathbf{J}^{R}_{1\rightarrow 0}) e^{\mathbf{L}\tau} (\mathbf{J}^{L}_{0\rightarrow 1}+\mathbf{J}^{L}_{1\rightarrow 0})\rangle \! \rangle}{\langle \! \langle \mathbf{J}^{R}_{0\rightarrow 1}+\mathbf{J}^{R}_{1\rightarrow 0} \rangle\! \rangle \langle\!\langle \mathbf{J}^{L}_{0\rightarrow 1}+\mathbf{J}^{L}_{1\rightarrow 0} \rangle\!\rangle }\theta(\tau) +\frac{\langle \! \langle (\mathbf{J}^{L}_{0\rightarrow 1}+\mathbf{J}^{L}_{1\rightarrow 0}) e^{-\mathbf{L}\tau} (\mathbf{J}^{R}_{0\rightarrow 1}+\mathbf{J}^{R}_{1\rightarrow 0})\rangle \! \rangle}{\langle \! \langle \mathbf{J}^{R}_{0\rightarrow 1}+\mathbf{J}^{R}_{1\rightarrow 0} \rangle\! \rangle \langle\!\langle \mathbf{J}^{L}_{0\rightarrow 1}+\mathbf{J}^{L}_{1\rightarrow 0} \rangle\!\rangle }\theta(-\tau) \\
&+ \frac{\langle\!\langle \mathbf{J}^\text{CAR}_\text{in}\rangle\!\rangle+\langle\!\langle \mathbf{J}^\text{CAR}_\text{out}\rangle\!\rangle+\langle\!\langle \mathbf{J}^\text{EC}_{L\rightarrow R}\rangle\!\rangle+\langle\!\langle \mathbf{J}^\text{EC}_{R\rightarrow L}\rangle\!\rangle}{\langle \! \langle \mathbf{J}^{R}_{0\rightarrow 1}+\mathbf{J}^{R}_{1\rightarrow 0} \rangle\! \rangle \langle\!\langle \mathbf{J}^{L}_{0\rightarrow 1}+\mathbf{J}^{L}_{1\rightarrow 0} \rangle\!\rangle }\delta(\tau) \approx 1+ \frac{\langle\!\langle \mathbf{J}^\text{CAR}_\text{in}\rangle\!\rangle+\langle\!\langle \mathbf{J}^\text{CAR}_\text{out}\rangle\!\rangle+\langle\!\langle \mathbf{J}^\text{EC}_{L\rightarrow R}\rangle\!\rangle+\langle\!\langle \mathbf{J}^\text{EC}_{R\rightarrow L}\rangle\!\rangle}{\langle \! \langle \mathbf{J}^{R}_{0\rightarrow 1}+\mathbf{J}^{R}_{1\rightarrow 0} \rangle\! \rangle \langle\!\langle \mathbf{J}^{L}_{0\rightarrow 1}+\mathbf{J}^{L}_{1\rightarrow 0} \rangle\!\rangle }\delta(\tau),
\end{split}
\label{Bare g2}
\end{equation}
where we in the last step have omitted the weak correlations between separate processes taking place on different islands. For the instantaneous correlations, the timing jitter of the detectors plays an important role, leading to a broadening of the delta peak. To account for the timing jitter, we convolve the correlation function with a Gaussian distribution of width $\sigma_D$, which is computed from a weighted average of the timing noise of the different processes
\begin{eqnarray}
\nonumber
\sigma_D &=& \frac{\langle\!\langle \mathbf{J}^\text{CAR}_\text{in}\rangle\!\rangle
\:\sigma^\text{CAR}_\text{in}
+\langle\!\langle \mathbf{J}^\text{EC}_{R\rightarrow L}\rangle\!\rangle
\:\sigma^\text{EC}_{R\rightarrow L}
+\langle\!\langle \mathbf{J}^\text{EC}_{L\rightarrow R}\rangle\!\rangle
\:\sigma^\text{EC}_{L\rightarrow R}
+\langle\!\langle \mathbf{J}^\text{CAR}_\text{out}\rangle\!\rangle
\:\sigma^\text{CAR}_\text{out}
}
{\langle\!\langle \mathbf{J}^\text{CAR}_\text{in}\rangle\!\rangle+\langle\!\langle \mathbf{J}^\text{EC}_{R\rightarrow L}\rangle\!\rangle+\langle\!\langle \mathbf{J}^\text{EC}_{L\rightarrow R}\rangle\!\rangle+\langle\!\langle \mathbf{J}^\text{CAR}_\text{out}\rangle\!\rangle}\\
&=& 460 \text{ $\mu$s}\edit{,}\label{eq:broad1}
\end{eqnarray}
where 
$\sigma^\text{CAR}_\text{in} = \sqrt{{\sigma_{\tau_{\mathrm{j}}(1\rightarrow 0)}^{L}}^{2} +{\sigma_{\tau_{\mathrm{j}}(1\rightarrow 0)}^{R}}^{2}}$, %
$\sigma^\text{EC}_{R\rightarrow L} = \sqrt{{\sigma_{\tau_{\mathrm{j}}(0\rightarrow 1)}^{L}}^{2} +{\sigma_{\tau_{\mathrm{j}}(1\rightarrow 0)}^{R}}^{2}}$,
$\sigma^\text{EC}_{L\rightarrow R} =
\sqrt{{\sigma_{\tau_{\mathrm{j}}(1\rightarrow 0)}^{L}}^{2} +{\sigma_{\tau_{\mathrm{j}}(0\rightarrow 1)}^{R}}^{2}}$ and 
$\sigma^\text{CAR}_\text{out} =
\sqrt{{\sigma_{\tau_{\mathrm{j}}(0\rightarrow 1)}^{L}}^{2} +{\sigma_{\tau_{\mathrm{j}}(0\rightarrow 1)}^{R}}^{2}}$ are the sums of the relevant jitter noises for each process. The superscripts $L$ and $R$ on the timing jitters $\sigma_{\tau_{\mathrm{j}}(i\rightarrow f)}$ here denote the left and right detector respectively. We then obtain the final expression
\begin{equation}
g^{(2)}_x(\tau) = 1+\frac{\alpha_2}{\sqrt{2\pi}\sigma_D} e^{-\tau^2/(2\sigma^2_D)},
\end{equation}
with
\begin{equation}
    \alpha_2 \equiv \frac{\langle\!\langle \mathbf{J}^\text{CAR}_\text{in}\rangle\!\rangle+\langle\!\langle \mathbf{J}^\text{CAR}_\text{out}\rangle\!\rangle+\langle\!\langle \mathbf{J}^\text{EC}_{L\rightarrow R}\rangle\!\rangle+\langle\!\langle \mathbf{J}^\text{EC}_{R\rightarrow L}\rangle\!\rangle}{\langle \! \langle \mathbf{J}^{R}_{0\rightarrow 1}+\mathbf{J}^{R}_{1\rightarrow 0} \rangle\! \rangle \langle\!\langle \mathbf{J}^{L}_{0\rightarrow 1}+\mathbf{J}^{L}_{1\rightarrow 0} \rangle\!\rangle } = 210~\text{ms},
\end{equation}
which is Equation~(2) of the main text. Here we see that the crossed Andreev reflections, together with elastic cotunneling, determine the integral over the distribution.

\subsection{B. Waiting time distributions}
We define the distribution of waiting times between two types of transitions $\alpha$ and $\beta$ as  \cite{PhysRevA.39.1200}
\begin{equation}
\mathcal{W}(\tau|t) \equiv \tilde P_{\beta \alpha}(t+\tau|t),
\label{WTD def}
\end{equation}
where $\tilde P_{\beta \alpha}(t+\tau|t)$ is the \emph{exclusive} probability density of observing a transition $\beta$ at time $t+\tau$ provided that a transition $\alpha$ occurred at time $t$, with no other transition $\beta$ taking place in between. For a steady state, the waiting time distribution depends only on the time difference, i.e., $\mathcal{W}(\tau|t) \equiv \mathcal{W}(\tau)$. In that case, we may express Equation~\eqref{WTD def} in terms of jump operators as
\begin{equation}
\mathcal{W}(\tau)=\frac{\langle\!\langle \mathbf{J}_\beta e^{(\mathbf{L}-\mathbf{J}_\beta)\tau} (\mathbf{J}_\alpha-\mathbf{J}_{\alpha \beta}) \rangle \!\rangle}{\langle \! \langle \mathbf{J}_\alpha \rangle \!\rangle }+\eta_0\delta(\tau),
\label{WTD def}
\end{equation}
with $\eta_0 \equiv \langle\!\langle \mathbf{J}_{\alpha \beta} \rangle \!\rangle/\langle \! \langle \mathbf{J}_\alpha \rangle \!\rangle$, which includes the instantaneous correlations of two-electron processes involving both $\alpha$ and $\beta$ transitions. Similar to the $g^{(2)}$-function, the instantaneous correlations are sensitive to the timing jitter of the detectors, as discussed for the cross-waiting time distribution below. The waiting time distribution is normalized, $\int_0^\infty \mathcal{W}(\tau)d\tau = 1$, and has the units of inverse time. Here, the delta function is defined such that $\int_0^\infty d\tau \delta(\tau) = 1$.
	
\subsubsection{1. Equation (3) in the main text}
We consider the distribution of waiting times between subsequent electrons tunneling out of the right island into the superconductor. To obtain this auto-waiting time distribution of the right island, we set $\mathbf{J}_\alpha = \mathbf{J}_\beta =\mathbf{J}^R_{1\rightarrow 0}$ and $\mathbf{J}_{\alpha \beta} = 0$ in Equation~\eqref{WTD def}, yielding the expression
	\begin{equation}
\mathcal{W}(\tau)=\frac{\langle\!\langle\mathbf{J}^R_{1\rightarrow 0}e^{(\mathbf{L}-\mathbf{J}^R_{1\rightarrow 0})\tau} \mathbf{J}^R_{1\rightarrow 0} \rangle \!\rangle}{\langle \! \langle\mathbf{J}^R_{1\rightarrow 0} \rangle \!\rangle }.
\label{WTD auto case}
\end{equation}
Similar to the auto-correlation $g^{(2)}$-function, we may use the fact that the two-electron processes involving also the left island are slow compared to the right island, and that the latter may therefore be treated as an effectively separate four-level system, whose dynamics is governed by the Liouvillian $\textbf{L}_R$ in Equation~\eqref{L4}. Treating the high-energy charge states $-1$ and $+2$ as short-lived virtual states, we find
\begin{equation}
\mathcal{W}(\tau) = \frac{1}{\langle \tau \rangle u }\left(e^{-\gamma(1-u) \tau/2}-e^{-\gamma(1+u)\tau/2}\right),
\label{Auto-WTD}
\end{equation}
which is Eq.~(3) in the main text. Here $\gamma$ is the inverse correlation time in Equation~\eqref{Effective rates auto-g2}, $\langle \tau \rangle = 1.2$~s is the average waiting time between two subsequent $1\rightarrow 0$ transitions on the right island, and $u=\sqrt{1-4/(\gamma \langle \tau \rangle)}$. Numerical calculations show that this approximation agrees very well with what we obtain by numerically evaluating Equation~\eqref{WTD auto case}.

\subsubsection{2. Equation (4) in the main text}
Last, we consider the distribution of waiting times between electrons tunneling out of the left island and electrons tunneling out of the right island. To obtain this cross-waiting time distribution, we set $\mathbf{J}_\alpha = \mathbf{J}^{L}_{1\rightarrow 0}$, $\mathbf{J}_\beta =\mathbf{J}^R_{1\rightarrow 0}$ and $\mathbf{J}_{\alpha \beta} =\mathbf{J}^\text{CAR}_\text{in}$ in Equation~\eqref{WTD def}, yielding
\begin{equation}
\mathcal{W}_{x}(\tau)=\frac{\langle\!\langle \mathbf{J}^R_{1\rightarrow 0} e^{(\mathbf{L}-\mathbf{J}^R_{1\rightarrow 0})\tau} (\mathbf{J}^L_{1\rightarrow 0}-\mathbf{J}_\text{in}^\text{CAR}) \rangle \!\rangle}{\langle \! \langle \mathbf{J}_{1\rightarrow 0}^L \rangle \!\rangle }+\eta_0 \delta(\tau) \equiv (1-\eta_0)\mathcal{W}_0(\tau) + \eta_0 \delta(\tau),
\label{expression for cross-WTD}
\end{equation}
where $\eta_0 = \langle \! \langle \mathbf{J}_\text{in}^\text{CAR} \rangle \!\rangle/\langle \! \langle \mathbf{J}_{1\rightarrow 0}^L \rangle \!\rangle =0.36$ and $\mathbf{J}^\text{CAR}_\text{in}$ is the superoperator of Cooper pair formation. A simple analytic expression for the first term, $\mathcal{W}_0(\tau)$, is challenging to obtain, however, it decays approximately as $\exp\left[-\gamma(1-u) \tau/2\right]$ for long waiting times. Similar to the cross-correlation $g^{(2)}$-function, the instantaneous correlations in the second term are broadened by the timing jitter of the detectors, resulting in a Gaussian distribution of width $\sigma_D =  \sqrt{(\sigma^L_{\tau_{\mathrm{j}}(1\rightarrow 0)})^2+(\sigma^R_{\tau_{\mathrm{j}}(1\rightarrow 0)})^2} = 410$ $\mu$s, here corresponding to the detector timing jitter of the $1\rightarrow 0$ transitions only. However, in contrast to the $g^{(2)}$-function, the waiting time distribution is sensitive to the time ordering of the events. In half of the Cooper pair formation processes, the timing jitter will cause the transition $1\rightarrow0$ on the right island to be detected \emph{before} the transition on the left. In those cases, the observed waiting time will extend until the next tunneling event, which to a good approximation is described by the auto-waiting time distribution in Equation~\eqref{Auto-WTD}. In conclusion, we thus find
\begin{equation}
\mathcal{W}_{x}(\tau)= (1-\eta_0) \mathcal{W}_0(\tau)+\frac{\eta_0}{2} \left[  \sqrt{\frac{2}{\pi}}\frac{1}{\sigma_D}\exp[-\tau^2/(2\sigma_D^2)]+ \mathcal{W}(\tau)\right],
\end{equation}
which is Equation~(4) in the main text. Just as for the cross-correlations, $g_x^{(2)}$, we see that the crossed Andreev reflections are important for the waiting time distribution, in particular, by giving rise to a large peak at short times.

\bibliography{supplement}